# Renormalization Group and Fermi Liquid Theory

A.C.Hewson

Dept. of Mathematics, Imperial College, London SW7 2BZ.

**Abstract**

We give a Hamiltonian based interpretation of microscopic Fermi liquid theory within a renormalization group framework. We identify the fixed point Hamiltonian of Fermi liquid theory, with the leading order corrections, and show that this Hamiltonian in mean field theory gives the Landau phenomenological theory. A renormalized perturbation theory is developed for calculations beyond the Fermi liquid regime. We also briefly discuss the breakdown of Fermi liquid theory as it occurs in the Luttinger model, and the infinite dimensional Hubbard model at the Mott transition.

email address: a.hewson@ic.ac.uk

# 1  Introduction

The recent theoretical attempts to understand the anomalous behaviour of the high temperature superconductors have raised some fundamental questions as to the form of the low energy excitations in strongly correlated low dimensional Fermi systems. Before the discovery of the high $T_c$ materials the superconductivities in most superconducting metals and compounds have been explained within the BCS theory as instabilities within a Fermi liquid induced by an attractive retarded interaction due to coupling with the phonons. Possible exceptions are the heavy fermion superconductors, such as $UPt_3$, where purely electronic mechanisms have been proposed for the effective interelectron attraction. Even in these cases the superconducting instability is believed to be one within a Fermi liquid. However, in their normal state high temperature superconductors are not good metals and their behaviour appears to differ from that of a conventional Fermi liquid (for a review of the experiments on these systems see [1]). This has led to conjectures that the normal state has no well defined quasi-particles at the Fermi level so that it can not be a Fermi liquid; there have been conjectures that the behaviour is better described by some form of 'marginal Fermi liquid' [2] or 'Luttinger liquid' [3]. More extreme breakdowns of Fermi liquid theory have also been proposed where no Fermi surface remains in the usual sense because the imaginary part of the self-energy is always finite [4]. These are still a controversial issues. However, these theories all propose that the superconductivity in the high $T_c$ materials is not an instability in a Fermi liquid and that we are dealing with a novel situation. As the characteristic feature of these materials is the $CuO_2$ plane, the basic question is whether the Fermi liquid breaks down in these two dimensional systems due to strong correlations induced between the d electrons at the $Cu$ sites. In one dimension it is well established that no matter how weak the inter-electronic interaction the Fermi liquid theory breaks down and the low energy excitations for short range repulsive interactions are well described by a suitably parametrized Luttinger liquid. Anderson's theory [3] for the high $T_c$ materials is based on the assumption that a similar Luttinger liquid state occurs in strongly correlated two dimensional models. If this is so, then could such a breakdown occur for higher dimensional systems with strong correlation? Is there a critical interaction strength for the breakdown of Fermi liquid theory for the two dimensional systems?

These are the sort of questions have led to a re-examination of Fermi liquid theory. It is nearly forty years since Landau proposed his phenomenological Fermi liquid theory [5], and it was very soon after that the approach was verified for microscopic models within the framework of many-body perturbation theory [6][8][7]. Since that time new techniques for tackling problems in condensed matter physics have been devised, such as the renormalization group approach, and others, such as 'bosonization', have been developed further. These approaches can bring a fresh perspective to the subject. Whether any of these techniques can provide answers to the question of the behaviour of two dimensional Fermi systems cannot be answered at this stage. What is clear, however, is that they can give new insights which are of value in themselves, and provide a new context for addressing these problems.

The bosonization approach in one dimension gives the essential physics for many interacting fermion models in a very simple way. This was shown in the pioneering work of Tomonaga [9], Luttinger [10], Mattis and Lieb [11], then further developed by Luther [12], Emery [13], Haldane [14] amongst others. Haldane recognized universal features in these results and showed that low energy behaviour of a large class of one dimensional interacting fermion systems can be derived from a suitably generalized Luttinger model. He coined the term 'Luttinger liquid' to describe these in analogy with concept of a Fermi liquid for higher dimensional Fermi systems. More recent bosonization work [15][16] has been concerned with showing how this approach can be generalized to higher dimensions, and showing how the Fermi liquid theory can emerge from bosons when they are constrained to the region of the Fermi surface. In this article, however, we will focus our attention on the renormalization group approach, both as developed by Wilson [17] in the 70s to tackle problems of critical phenomena but also as originally developed in field theory, QED etc, as a reorganization of perturbation theory. Recently the Wilson renormalization approach to these fermion models has been the subject of a review by Shankar [18]. In this article we cover similar ground but with a different emphasis and different examples so that there is in detail little overlap between the two surveys. The basic message, however, I think is the same, that the renormalization group gives us new insights into Fermi liquid theory, its stability and the circumstances which may cause it to breakdown. We develop a renormalized perturbation theory which can be directly related to the Wilson type of calculations, and also to the microscopic derivation of Fermi liquid theory.

Before embarking on the renormalization group approach we briefly look at the main features of Fermi liquid theory as originally introduced by Landau. The Landau phenomenological theory [5] is based on the assumption that the single (quasi)particle excitations at very low temperatures of an interacting Fermi system are in one-to-one correspondence with those of the non-interacting system. In terms of the one electron states a total energy functional $E_{\rm tot}$ is constructed of the form,

$$E_{\rm tot} = E_{\rm gs} + \sum_{\alpha,\sigma} \tilde{\epsilon}^{(0)}_{\alpha,\sigma} \delta n_{\alpha,\sigma} + \frac{1}{2} \sum_{\alpha\alpha',\sigma\sigma'} f_{\sigma,\sigma'}(\alpha,\alpha') \delta n_{\alpha,\sigma} \delta n_{\alpha',\sigma'} + ... \qquad (1)$$

where $E_{\rm gs}$ is the ground state energy, $\delta n_{\alpha,\sigma}$ is the deviation in occupation number of the single particle state $|\alpha\rangle, \sigma$ with an excitation energy $\tilde{\epsilon}^{(0)}_{\alpha,\sigma}$ from its ground state value, and $f^{\sigma,\sigma'}_{\alpha,\alpha'}$ is the leading term due to the quasiparticle interactions. A free energy functional $F$ is constructed from (1), retaining only the first two terms, together with the Fermi–Dirac form for the entropy of the quasiparticles. Minimization of $F$ with respect to $\delta n_\alpha$ leads to asymptotically exact results for the thermodynamic behaviour as $T, H \to 0$ for systems in a normal paramagnetic ground state. The reason why the higher order terms in (1) give negligible effects in this limit is because the expectation value of $\delta n_\alpha$, $\langle \delta n_\alpha \rangle \to 0$ as $T \to 0$ and $H \to 0$. The effective quasiparticle energy $\tilde{\epsilon}_{\alpha,\sigma}$ in the presence of other excitations is given by

$$\tilde{\epsilon}_{\alpha,\sigma} = \tilde{\epsilon}_{\alpha,\sigma}^{(0)} + \sum_{\alpha',\sigma'} f_{\sigma,\sigma'}(\alpha,\alpha')\langle \delta n_{\alpha',\sigma'}\rangle. \tag{2}$$

From this result asymptotically exact results for the specific, susceptibility and other low temperature properties can be deduced. By taking into account scattering of the quasiparticles the low temperature behaviour of the transport coefficients can be calculated, and equations for the coherent excitation of quasiparticle-quasihole pairs lead to a description of collective modes such as zero sound. Application and further discussion of the Landau phenomenological approach can be found in references [19][20]. This approach was verified within the framework of many-body perturbation theory in the early 60s and detailed derivations of the basic theory can be found in the papers of Luttinger [6] and the books of Nozières [7], and Abrikosov, Gorkov and Dzyaloshinskii [8]. Due to the mathematical complexity of the diagrammatic perturbation theory, however, some of the more intuitive ideas of Landau are lost. Here we intend to show that the renormalization group approach can make a useful conceptual link between the two approaches. For those not familiar with the renormalization group we very brief review of the philosophy of this approach as developed by Wilson for tackling the problems of critical phenomena, and also for the Kondo problem [17].

The renormalization group is a mapping $R$ of a Hamiltonian $H(\mathbf{K})$, which is specified by a set of interaction parameters or couplings $\mathbf{K} = (K_1, K_2, \ldots)$ into another Hamiltonian of the same form with a new set of coupling parameters $\mathbf{K}' = (K_1', K_2', \ldots)$. This is expressed formally by

$$R\{H(\mathbf{K})\} = H(\mathbf{K}'), \tag{3}$$

or equivalently,

$$R(\mathbf{K}) = \mathbf{K}', \tag{4}$$

where the transformation is in general non-linear. In applications to critical phenomena the new Hamiltonian is obtained by removing short range fluctuations to generate an effective Hamiltonian valid over larger length scales. In the cases we consider here the transformations are generated by eliminating higher energy excited states to give a new effective Hamiltonian for the reduced energy scale for the lower lying states. The transformation is characterized by a parameter, say $\alpha$, which specifies the ratio of the new length or energy scale to the old one such that a sequence of transformations, generates a sequence of points or, where $\alpha$ is a continuous variable, a trajectory in the parameter space $\mathbf{K}$. The transformation is constructed so that it satisfies

$$R_{\alpha'}\{R_\alpha(\mathbf{K})\} = R_{\alpha+\alpha'}(\mathbf{K}). \tag{5}$$

The key concept of the renormalization group is that of a fixed point, a point $\mathbf{K}^*$ which is invariant under the transformation,

$$R_\alpha(\mathbf{K}^*) = \mathbf{K}^*. \tag{6}$$

The trajectories generated by the repeated application of the renormalization group tend to be drawn towards, or expelled from, the fixed points. The behaviour of the trajectories near a fixed point can usually be determined by linearizing the transformation in the neighbourhood of the fixed point. If in the neighbourhood of a particular fixed point $\mathbf{K} = \mathbf{K}^* + \delta \mathbf{K}$ then, expanding $R_\alpha(\mathbf{K})$ in powers of $\delta \mathbf{K}$,

$$R_\alpha(\mathbf{K}^* + \delta \mathbf{K}) = \mathbf{K}^* + \mathbf{L}_\alpha^* \delta \mathbf{K} + \mathrm{O}(\delta \mathbf{K}^2), \tag{7}$$

where $\mathbf{L}_\alpha^*$ is a linear transformation. If the eigenvectors and eigenvalues of $\mathbf{L}_\alpha^*$ are $\mathbf{O}_n^*(\alpha)$ and $\lambda_n^*(\alpha)$, and if these are complete then they can be used as a basis for a representation of the vector $\delta \mathbf{K}$,

$$\delta \mathbf{K} = \sum_n \delta K_n \mathbf{O}_n^*(\alpha), \tag{8}$$

where $\delta K_n$ are the components. How the trajectories move in the region of a particular fixed point depends on the eigenvalues $\lambda_n^*$. if we act $m$ times on a point $\mathbf{K}$ in the neighbourhood of a fixed point $\mathbf{K}^*$ then, from (8), we find

$$R_\alpha^m(\mathbf{K}^* + \delta \mathbf{K}) = \mathbf{K}^* + \sum_n \delta K_n \lambda_n^{*m} \mathbf{O}_n^*(\alpha), \tag{9}$$

provided all the points generated by the transformation are in the vicinity of the fixed point so that the linear approximation (9) remains valid. Eigenvalues with $\lambda_n^* > 1$ are termed relevant and the corresponding components of $\delta K_n$ in (9) increase with $m$ while those with eigenvalues $\lambda_n^* < 1$, termed irrelevant, get smaller with $m$. Those with $\lambda_n^* = 1$ to linear order do not vary with $m$ and are marginal. The eigenvalues of the linearized equation lead to a classification of the fixed points: stable fixed points have only irrelevant eigenvalues so $\delta \mathbf{K} \to 0$, unstable fixed points have one or more relevant eigenvalues and the trajectories are eventually driven away from the fixed point. A marginal fixed point has no relevant eigenvalues and at least one marginal one. Whether a trajectory is ultimately driven towards or away from the fixed point in this case depends on the non-linear corrections.

The application of the renormalization group to critical phenomena is based on the fact that at a second order phase transition the correlation length becomes infinite so that at the critical temperature the system appears the same on all length scales. As a result at the critical point the parameter changes on applying the renormalization group transformation at the critical point are small and tend to zero as the length scale becomes large compared to the lattice spacing. This behaviour corresponds to the asymptotic approach of the trajectory to a fixed point of the renormalization group transformation. For small deviations away from the critical temperature, this scale invariance no longer holds and under the renormalization group transformation the system trajectory is eventually driven away from the fixed point. For this to occur the fixed point must be an unstable one. For critical behaviour of the form $|T - T|^\nu$ the critical exponent $\nu$ can be calculated from the relevant eigenvalues of the linearized renormalization group equations about this fixed point (for further details see for example [21]). The universality of the critical exponents follows when a whole class of systems scale to the same fixed point, and consequently all have the same critical behaviour.

In the situations we consider here the renormalization group transformations are associated with the elimination of higher energy states to generate an effective Hamiltonian for the lower lying levels. If the system does not undergo a phase transition, or develop an energy gap, then we expect the physics on the lowest energy scales not to change significantly so that we find a limiting form for the effective Hamiltonian as the energy scale is reduced to zero. This Hamiltonian must correspond to a stable or marginal fixed point of the renormalization group transformation. It will be this fixed point Hamiltonian, and the leading correction terms, which determines the lowest lying excitations of the system and hence the thermodynamic behaviour as $T \to 0$.

As one universal form of low energy behaviour is that of a Fermi liquid it seems a natural question to ask whether we can reformulate Fermi liquid theory within the renormalization group framework. The idea of a Fermi liquid fixed point has been used frequently in condensed matter theory (see for example [22]) but very little work has been done in clarifying the concept and relating it to existing phenomenological and microscopic theory. Renormalization group calculations for fermion systems, in which higher energy excitations are eliminated and an effective Hamiltonian obtained for the lowest energy scales, are difficult to carry out. There have been a some calculations for translationally invariant systems and lattice models (see [23][18] and references therein) but the most successful calculations of this type have been those for magnetic impurity models [17][24], the Kondo and Anderson models for a magnetic 3d transition or 4f rare earth ion in a metallic host. These are relatively simple systems yet with non-trivial physics, with a local moment regime at higher temperatures and a crossover to Fermi liquid behaviour at low temperatures and eventually to a spin compensated ground state. Apart from the extensive renormalization group calculations there are perturbational results [25] as well as an exact solutions for the thermodynamics [26] for these models. It will be instructive to look at the renormalization group results for these systems for which we have a rather complete quantitative description from a range of theoretical approaches. The insights gained from these specific examples will enable us to make conjectures, and devise methods, to apply to a very general class of systems.

## 2   Fermi liquid theory of impurity models

We base our discussion primarily on the Anderson model [27] as it gives a more general description of a magnetic impurity than the Kondo model because it includes the possibility of charge fluctuations at the impurity site. In its simplest form the model has an impurity d (f) level $\epsilon_d$, taken to be non-degenerate, which is hybridized with the host conduction electrons via a matrix element $V_{\mathbf{k}}$. When the interaction term $U$ between the electrons in the local d state is included the Hamiltonian has the form,

$$H = \sum_\sigma \epsilon_{d,\sigma} c^\dagger_{d,\sigma} c_{d,\sigma} + U n_{d,\uparrow} n_{d,\downarrow} + \sum_{\mathbf{k},\sigma}(V_\mathbf{k} c^\dagger_{d,\sigma} c_{\mathbf{k},\sigma} + V^*_\mathbf{k} c^\dagger_{\mathbf{k},\sigma} c_{d,\sigma}) + \sum_{\mathbf{k},\sigma} \epsilon_{\mathbf{k},\sigma} c^\dagger_{\mathbf{k},\sigma} c_{\mathbf{k},\sigma}, \quad (10)$$

where $\Delta(\omega) = \pi \sum_\mathbf{k} |V_\mathbf{k}|^2 \delta(\omega - \epsilon_\mathbf{k})$ is the function which controls the width of the virtual bound state resonance at $\epsilon_d$ in the non-interacting model ($U = 0$). In the limit of a wide conduction band with a flat density of states $\Delta(\omega)$ becomes independent of $\omega$ and can be taken as a constant $\Delta$. For $\epsilon_d \ll \epsilon_\mathrm{F}$, $\epsilon_d + U \gg \epsilon_\mathrm{F}$, where $\epsilon_\mathrm{F}$ is the Fermi level and $|\epsilon_d - \epsilon_\mathrm{F}|, |\epsilon_d + U - \epsilon_\mathrm{F}| \gg \Delta$, the model is equivalent to the Kondo model with an antiferromagnetic interaction $J\rho_0 = U\Delta/\pi|\epsilon_d - \epsilon_\mathrm{F}||\epsilon_d + U - \epsilon_\mathrm{F}|$, where $\rho_0$ is the conduction electron density of states.

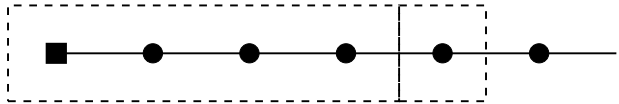

Figure 1. A linear chain form of the Anderson impurity model with the impurity at one end (filled square) coupled by the hybridization to a tight-binding chain of conduction states (filled circles). Iterative diagonalization proceeds by diagonalizing a finite chain, adding a further conduction electron site to the chain, and repeating the process.

We look first of all at the renormalization group calculations of Wilson who was the first to obtain a precise description of the low temperature behaviour of the model in the Kondo regime where the spin fluctuation effects dominate. We need not go into the technical details of the numerical renormalization group method he devised but it will help to have a brief overview to see what was involved. The model was cast in the form with the conduction electron states in the form of a tight-binding chain with the impurity at one end coupled via the hybridization (see figure 1). By iterative diagonalization of chains of increasing length, in which a finite number of the low lying states were retained at each stage, the lowest excitations on a decreasing energy scale were obtained (the details can be found in the original papers of Wilson [17], and Krishnamurthy, Wilkins and Wilson [24]). The transformation that maps the model on one energy scale to one on a reduced energy scale, when suitably scaled, is a renormalization group transformation. The higher energy excitations, which are eliminated at each stage, modify the couplings of the model. In general further interaction terms, not present in the original model, are introduced. It can be shown that the effective Hamiltonian for the lowest energy scales found by this technique can be expressed in the same form as the original Anderson model but with modified (renormalized) parameters, [28][29], $\tilde{\Delta} = \pi \sum_\mathbf{k} |\tilde{V}_\mathbf{k}|^2 \delta(\omega - \epsilon_\mathbf{k})$, $\tilde{U}$, and $\tilde{\epsilon}_d$, corresponding to the effective Hamiltonian,

$$H_\mathrm{eff} = \sum_{\mathbf{k},\sigma} \epsilon_{\mathbf{k},\sigma} c^\dagger_{\mathbf{k},\sigma} c_{\mathbf{k},\sigma} + \sum_{\mathbf{k},\sigma}(\tilde{V}_\mathbf{k} \tilde{c}^\dagger_{d,\sigma} c_{\mathbf{k},\sigma} + \tilde{V}^*_\mathbf{k} c^\dagger_{\mathbf{k},\sigma} \tilde{c}_{d,\sigma}) + \sum_\sigma \tilde{\epsilon}_{d,\sigma} \tilde{c}^\dagger_{d,\sigma} \tilde{c}_{d,\sigma} + \tilde{U} \tilde{n}_{d,\uparrow} \tilde{n}_{d,\downarrow}, \quad (11)$$

where the energy level $\tilde\epsilon_{d,\sigma}$ is measured with respect to the Fermi level.

For the local moment regime of the model there is only one energy scale involved which is the Kondo temperature $T_{\rm K}$. The important point which makes the renormalized effective Hamiltonian (11) tractable at low temperatures is that it describes excitations from the *exact ground state* so that the interaction term $\tilde U$ between excitations only comes into play when there is more than one excitation from the ground state. From the renormalization group perspective the interaction terms are the leading irrelevant corrections to the free fermion fixed point. They tend to zero under the rescaling of the renormalization group as the low energy fixed point is approached ($T \to 0$). If a single particle excitation or single hole excitation is created this interaction term plays no role and can be omitted and the one electron Hamiltonian remaining can be diagonalized and written in the form,

$$H_{\rm eff}(\tilde U = 0) = \sum_{l,\sigma} \tilde\epsilon^{(0)}_{l,\sigma} c^\dagger_{l,\sigma} c_{l,\sigma}, \qquad (12)$$

where $\tilde\epsilon^{(0)}_{l,\sigma}$ are the one-electron energies, and $c^\dagger_{l,\sigma}$ and $c_{l,\sigma}$ the corresponding creation and annihilation operators. This is the fixed point Hamiltonian as $T \to 0$, and corresponds to free or non-interacting fermions. Comparing this with the Landau theory we can identify the excitation energy $\tilde\epsilon^{(0)}_{l,\sigma}$ as the energy of the non-interacting quasiparticles in the free energy functional (1). Rather than diagonalize (11) explicitly for the excitation energies we can solve for the Green's function of the quasiparticles by the equation of motion technique which gives a closed set of equations for $\tilde U = 0$. This gives

$$\tilde G_{{\rm d}\sigma}(\omega) = \frac{1}{\omega - \tilde\epsilon_{\rm d} + i\tilde\Delta}. \qquad (13)$$

the quasiparticle energies correspond to the poles of this Green's function. The corresponding spectral density or quasiparticle density of states $\tilde\rho_d(\omega)$ can be straightforwardly be deduced and is given by

$$\tilde\rho_d(\omega) = \frac{1}{\pi} \frac{\tilde\Delta}{\left[(\omega - \tilde\epsilon_d)^2 + \tilde\Delta^2\right]}, \qquad (14)$$

corresponding to a Lorentzian resonance, the so-called Kondo resonance.

For $\tilde U \neq 0$ (11) can be written using the single particle eigenstates of (12) as a basis,

$$H_{\rm eff} = \sum_{l,\sigma} \epsilon_{l,\sigma} c^\dagger_{l,\sigma} c_{l,\sigma} + \tilde U \sum_{l,l'} \sum_{l'',l'''} \alpha^*_l \alpha_{l'} \alpha^*_{l''} \alpha_{l'''} c^\dagger_{l,\uparrow} c_{l',\uparrow} c^\dagger_{l'',\downarrow} c_{l''',\downarrow} \qquad (15)$$

where

$$\tilde c^\dagger_{d,\sigma} = \sum_l \alpha^*_{l,\sigma} c^\dagger_{l,\sigma}. \qquad (16)$$

As it is the excitations of the interacting system from its ground state that are described by the effective Hamiltonian (11), it is appropriate to transform this Hamiltonian to operators which describe the single particle excitations,

$$c^\dagger_{l,\sigma} = p^\dagger_{l,\sigma} \quad c_{l,\sigma} = p_{l,\sigma} \quad \epsilon_{l,\sigma} > \epsilon_{\rm F}, \qquad (17)$$

$$c_{l,\sigma}^{\dagger} = h_{l,\sigma} \quad c_{l,\sigma} = h_{l,\sigma}^{\dagger} \quad \epsilon_{l,\sigma} < \epsilon_{\rm F}, \tag{18}$$

where the ground state is such that $p_{l,\sigma}|0\rangle = 0$ and $h_{l,\sigma}|0\rangle = 0$. The interaction term has to be normal ordered in terms of these operators so that $H_{\rm eff}|0\rangle = 0$, and the Hamiltonian describes interactions *only between excitations from the ground state*. In the mean field approximation the expectation value of operators in the interaction terms such as $p_{l,\uparrow}^{\dagger} p_{l',\uparrow} p_{l'',\downarrow}^{\dagger} p_{l''',\downarrow}$ are approximated by

$$\langle p_{l,\uparrow}^{\dagger} p_{l',\uparrow} p_{l'',\downarrow}^{\dagger} p_{l''',\downarrow}\rangle = \langle p_{l,\uparrow}^{\dagger} p_{l,\uparrow}\rangle \langle p_{l'',\downarrow}^{\dagger} p_{l'',\downarrow}\rangle \delta_{l,l'} \delta_{l'',l'''}. \tag{19}$$

The quasiparticle energy of the Landau theory in the presence of other excitations can be identified as the effective one particle energy in this approximation,

$$\tilde{\epsilon}_{l,\sigma} = \tilde{\epsilon}_{l,\sigma}^{(0)} + \tilde{U}|\alpha_l|^2 \sum_{l'} |\alpha_{l'}|^2 \langle \delta n_{l',-\sigma}\rangle. \tag{20}$$

where $\langle \delta n_{l',\sigma}\rangle = \langle p_{l',\sigma}^{\dagger} p_{l',\sigma}\rangle$ for $\epsilon_{l'} > \epsilon_{\rm F}$ and $\delta n_{l',\sigma} = -\langle h_{l',\sigma}^{\dagger} h_{l',\sigma}\rangle$ for $\epsilon_{l'} < \epsilon_{\rm F}$. as $|\alpha_l|^2$ is proportional to $1/N_s$, where $N_s$ is the number of sites, because the scattering potential is due to a single impurity, the energy shift in (18) is of the order $1/N_s$.

As $\langle p_{l,\sigma}^{\dagger} p_{l,\sigma}\rangle \to 0$ as $T \to 0$ the quasiparticle interaction does not contribute to the linear term in the specific heat which can be calculated from the non-interacting quasiparticle Hamiltonian (12). Hence the impurity specific heat coefficient $\gamma_{\rm imp}$ is given by

$$\gamma_{\rm imp} = \frac{2\pi^2 k_{\rm B}^2}{3} \tilde{\rho}_d(\epsilon_{\rm F}). \tag{21}$$

where $\tilde{\rho}_d(\omega)$ is the impurity quasiparticle density of states defined in equation (14). in calculating the susceptibility the quasiparticle interaction has to be taken into account. Using the mean field approximation for this interaction given above the impurity susceptibility can be deduced and expressed in the form,

$$\chi_{\rm imp} = \frac{(g\mu_{\rm b})^2}{2}\tilde{\rho}_d(\epsilon_{\rm F})(1 + \tilde{U}\tilde{\rho}_d(\epsilon_{\rm F})) \tag{22}$$

for a flat wide conduction band. The total charge susceptibility at $T = 0$, $\chi_c = dn_0/d\epsilon_{\rm F}$, where $n_0$ is the expectation value of the total number operator for the electrons, can be calculated following precisely the same argument. The result for the impurity contribution is

$$\chi_{c,{\rm imp}} = 2\tilde{\rho}_d(\epsilon_{\rm F})(1 - \tilde{U}\tilde{\rho}_d(\epsilon_{\rm F})). \tag{23}$$

eliminating the term in $\tilde{U}$ between (22) and (23), and the term in $\tilde{\rho}_d(\epsilon_{\rm F})$ using (19), gives the well known Fermi liquid relation for the impurity model relating $\gamma_{\rm imp}$, $\chi_{\rm imp}$ and $\chi_{c,{\rm imp}}$,

$$\frac{4\chi_{\rm imp}}{(g\mu_{\rm B})^2} + \chi_{c,{\rm imp}} = \frac{6\gamma_{\rm imp}}{\pi^2 k_{\rm B}^2}. \tag{24}$$

The interesting physics of this model occurs in the strong correlation regime when $U$ is large and $\epsilon_d$ is well below the Fermi level so that the d electron at the impurity is localized and has an occupation number $n_d \sim 1$. This is the local moment regime in which the model maps into a Kondo model. At high temperatures $T \gg T_{\rm K}$, where $T_{\rm K}$ is the Kondo temperature, there is a Curie law susceptibility associated with the local moment but with logarithmic corrections, $\ln(T/T_{\rm K})$, which reduce the effective moment. In the low temperature regime (11) is valid for $T \ll T_{\rm K}$ when the impurity moment is fully screened giving a finite susceptibility at $T = 0$. The charge susceptibility on the other hand must tend to zero in the strong correlation limit as the $d$ electron at the impurity is localized. The condition $\chi_{\rm c,imp} = 0$ is achieved in the quasiparticle picture by the interaction term $\tilde{U}$, which self-consistently constrains the many-body resonance at the Fermi-level so that the impurity occupation is maintained with $n_d = 1$. This was first pointed out by Nozières [30]. This condition can be used to deduce $\tilde{U}$. In the case of particle-hole symmetry $\tilde{\epsilon}_d = 0$ and the resonance in the quasiparticle density of states (14) is peaked at the Fermi level and $\tilde{\rho}_d(\epsilon_{\rm F}) = 1/\pi\tilde{\Delta}$. We then find on substituting into equations (22) and (23),

$$\tilde{U} = \pi\tilde{\Delta} = 4T_{\rm K}, \qquad (25)$$

where the Kondo temperature, which is the only relevant energy scale for the model in the local moment regime, is defined by $\chi_{\rm imp} = (g\mu_{\rm B})^2/4T_{\rm K}$.

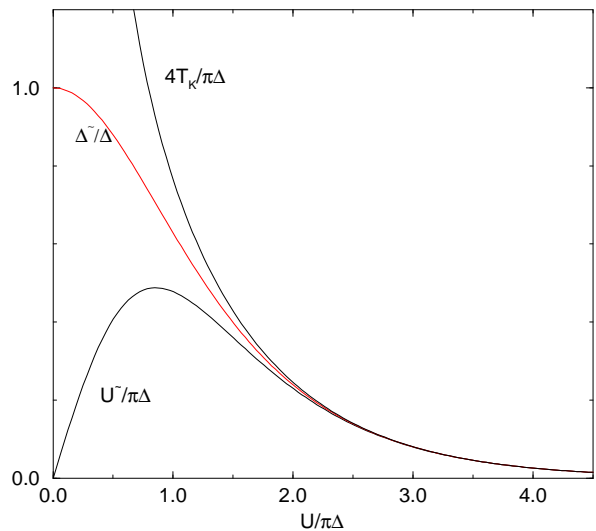

Figure 2 A plot of the renormalized parameters $\tilde{U}$ and $\tilde{\Delta}$ for the symmetric Anderson model in terms of the bare parameters $U$ and $\Delta$. In the comparison of these parameters with $4T_{\rm K}$ for $U \gg \pi\Delta$ the value of $T_{\rm K}$ is given by (27).

These parameters give for the '$\chi/\gamma$' or Wilson ratio $R$,

$$R = \frac{4\pi^2 k_B^2 \chi_{\rm imp}}{3(g\mu_B)^2 \gamma_{\rm imp}} = \frac{\chi_{\rm imp}/\chi_0}{\gamma_{\rm imp}/\gamma_0} = 2, \qquad (26)$$

and so is enhanced over the free electron value of unity, where $\chi_0$ and $\gamma_0$ are the susceptibilities and specific heat coefficients for the conduction electrons alone. The argument used here is essentially a reformulation of the one originally given by Nozières [30]. The result $R = 2$ was first derived by Wilson from his numerical renormalization group results [17].

As the Anderson model is integrable and exact solutions exist for the thermodynamic behaviour it is possible to deduce the renormalized parameters exactly over the full parameter regime. For the symmetric Anderson model the parameters $\tilde{U}$ and $\tilde{\Delta}$ have been calculated using the exact Bethe ansatz results [26] from equations (24) and (26). These are shown in figure 2 for the symmetric model ($n_d = 1$) plotted as a function of $U/\pi\Delta$. At weak coupling $\tilde{U} \sim U$ as one would expect and $\tilde{U}$ and $\tilde{\Delta}$ are independent energy scales. At strong coupling there is only the single energy scale $T_K$ given by

$$T_K = U \left(\frac{\Delta}{2U}\right)^{1/2} e^{-\pi U/8\Delta + \pi\Delta/2U}, \qquad (27)$$

with $\tilde{U}$ and $\tilde{\Delta}$ given by (25). Using these equations the form of the non-interacting quasiparticle density of states $\tilde{\rho}_d(\omega)$ given by (14), normalized by $1/\pi\tilde{\Delta}$, for various values of $U$ is shown in figure 3 illustrating the exponential narrowing of the resonance at the Fermi level in the Kondo regime ($U > \pi\Delta$) as $U$ is increased.

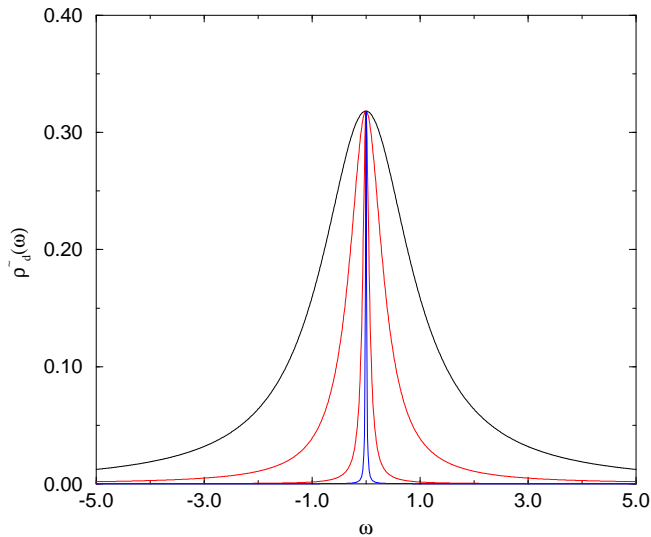

Figure 3. The quasi-particle density of states $\tilde{\rho}_d(\omega)$ normalized by $1/\pi\tilde{\Delta}$ as a function

of $U$ for $U/\pi$=0, 0.5, 1.0, 1.5 (in order of decreasing width) and $\Delta = 1.0$ for the particle-hole symmetric Anderson model.

This approach can be used for other magnetic impurity models, such as the $N$ fold degenerate Anderson model ($U = \infty$) and the $n = 2S$ multi-channel Kondo model [28]. Though explicit renormalization group calculations do not exist for most of these models one can conjecture the form of the quasiparticle Hamiltonian to describe the Fermi liquid regime and deduce the Fermi liquid relations. These relations can be checked with the exact results known from either the Bethe ansatz or microscopic Fermi liquid theory. For these models there are more renormalized parameters to take into account but in the strong correlation regime they can all be expressed in terms of the Kondo temperature. We give an example in figure 4 where we plot the quasiparticle density of states in the Kondo regime for the $N$ fold degenerate model for different values of $N$ [28]. The case $N = 2$ corresponds to the Kondo resonance at the Fermi level of the large $U$ Anderson model as shown in figure 3. With increasing $N$ this resonance narrows and moves towards $T_{\rm K}$ above the Fermi level, becoming a delta function at $\omega = T_{\rm K}$ in the large $N$ limit. Much of the low energy and low temperature physics of this model can be deduced from this asymmetric form for the quasiparticle density of states.

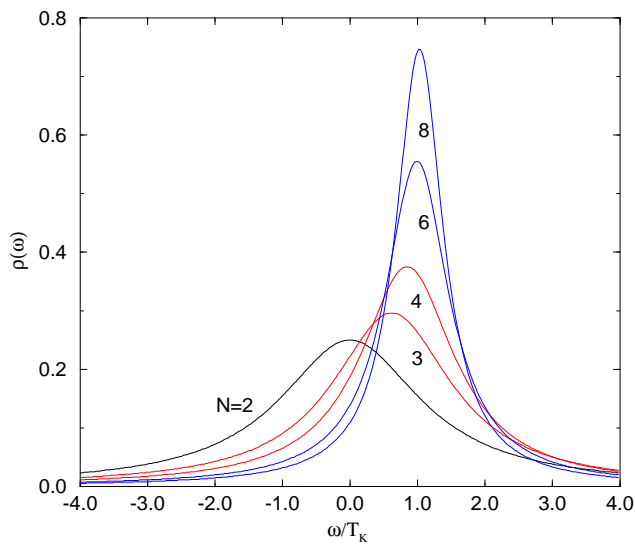

Figure 4. The quasiparticle density of states $\tilde{\rho}(\omega)$ for the N-fold degenerate Anderson model ($U = \infty$) in the Kondo regime for $N = 2, 3, 4, 6, 8$, as indicated, as a function of $\omega/T_{\rm K}$ where $T_{\rm K}$ is the Kondo temperature.

What general conclusions about Fermi liquid theory might we be tempted to draw from these explicit renormalization group calculations? First of all, there is strong parallel with the original phenomenological approach of Landau. The renormalization group calculations have given us an explicit Hamiltonian to describe the low energy excitations. The mean field expectation value of this Hamiltonian corresponds to the expectation value of the Landau total energy functional. The fact that this energy functional had to be expressed in terms of the deviations in the occupations of the states from their ground state values has its parallel in the renormalization group fixed point Hamiltonian, the Hamiltonian had to be normal ordered because it only describes the excitations from the ground state. The fact that the mean field theory is asymptotically exact as $T \to 0$ is because the deviation in the occupation of these states tends to zero in this limit; this corresponds to a low density of excitations. For the mean field treatment of the Hamiltonian to be valid there must be no singular scattering between the quasiparticles at low energies. Such singular scattering occurs in one dimensional models, and in higher dimensions with attractive interactions. We will discuss this more fully later.

A microscopic derivation of these Fermi liquid relations for the symmetric Anderson model has been given in a series of papers by Yamada and Yosida [25]. Their calculations were based on an expansion in powers of $U$ for the symmetric model. Though it is not possible to sum explicitly the dominant terms in the strong correlation limit in this model exact relations can be deduced corresponding to Fermi liquid theory in terms of the local self-energy $\Sigma(\omega)$ evaluated at the Fermi level ($\omega = 0$), its derivative and the irreducible four point vertex. Setting up a correspondence with the above we can identify the renormalized parameters, $\tilde{\epsilon}_\mathrm{d}$, $\tilde{\Delta}$ and $\tilde{U}$, in terms of these [29]. With $z$, the wavefunction renormalization factor, given by

$$z = \frac{1}{1 - \Sigma'(0)}, \qquad (28)$$

where prime denotes a derivative with respect to $\omega$ (evaluated at the Fermi level $\omega = 0$), the renormalized parameters are given by

$$\tilde{\epsilon}_{\mathrm{d},\sigma} = z(\epsilon_{\mathrm{d},\sigma} + \Sigma_\sigma(0,0)), \quad \tilde{\Delta} = z\Delta, \quad \tilde{U} = z^2 \Gamma_{\uparrow,\downarrow}(0,0), \qquad (29)$$

where $\Gamma_{\sigma,\sigma'}(\omega,\omega')$ is the irreducible four point vertex function. The self-energy has two arguments set to zero to indicate that it is evaluated not only for $\omega = 0$ but also at zero temperature and in zero field.

We have succeeded in showing that from the Wilson normalization group approach for the impurity model a quasiparticle Hamiltonian can be derived corresponding to a renormalized Anderson model, and that this in the mean field approximation gives the Landau total energy functional. Both the Landau and the renormalization group approaches give exact results as $T \to 0$. If we try to extend the renormalization group calculations beyond the Fermi liquid regime by generating an effective Hamiltonian over a higher energy range then further and more complicated interactions will come into play as the renormalization group trajectories are no longer under the exclusive influence of the low energy fixed point. On higher energy scales the effective Hamiltonian approach only becomes useful when the trajectories once again become dominated by a particular fixed point. For the crossover regime the effective Hamiltonian approach is not feasible and we have to rely on explicit numerical renormalization group calculations of the energy levels for predictions. The spectral density $\rho_d(\omega)$ of the d electron Green's function over the full energy range for the symmetric Anderson model, as calculated by the numerical renormalization group [31], is shown in figure 5. The quasiparticle density of states only describes the small region $\omega$ near the Fermi-level, the extra peaks at $\omega \sim \pm U/2$ are associated with the broadened 'atomic' peaks. The magnetic impurity problem is rather a special case in that we have a numerical renormalization results over all the relevant energy scales. As such calculations are difficult to carry out in general for other interacting fermion systems it is worth while to look at another renormalization procedure in which we make a perturbation expansion in terms of the fully dressed quasiparticles [32]. This corresponds to a reorganization of perturbation theory in close analogy with that developed originally for quantum electrodynamics where the expansion is in terms of the particles with their observed masses and charges (interaction strengths), ie. fully dressed. In the field theory case the reorganization of the perturbation theory was a necessary one to eliminate the ultraviolet divergences and obtain finite results. Due to the finite upper cut-offs in condensed matter physics problems such a reorganization of perturbation theory is not a necessary on one in order to get finite results, nevertheless, it makes a lot of sense to work with the fully dressed quasiparticles, particularly for systems where the renormalization effects are very large such as in heavy fermion systems. What is more it integrates the microscopic perturbation theory derivation of Fermi liquid theory with the more intuitive Landau approach.

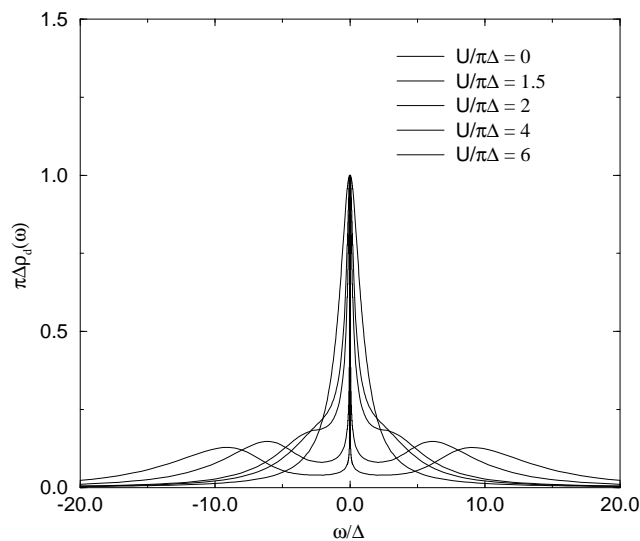

Figure 5. The spectral density $\pi\Delta\rho_d(\omega)$ for the d electron Green's function of the symmetric Anderson model for values of $U/\pi\Delta$ as indicated (see reference [31]).

## 3 A renormalized perturbation approach

We continue with the Anderson model and consider the retarded one particle double-time Green's function for the d-electron expressed in the form,

$$G_{d\sigma}(\omega) = \frac{1}{\omega - \epsilon_d - \Lambda(\omega) + i\Delta(\omega) - \Sigma_\sigma(\omega)}, \tag{30}$$

where $\Delta(\omega)$ was defined earlier and $\Lambda(\omega) = P\sum_{\mathbf{k}} |V_{\mathbf{k}}|^2/(\omega - \epsilon_{\mathbf{k}})$. We continue to work in the wide band limit where $\Delta(\omega)$ is independent of $\omega$ and $\Lambda(\omega) \to 0$. The function $\Sigma_\sigma(\omega)$ is the proper self-energy within a perturbation expansion in powers of the local interaction $U$. We will need the corresponding irreducible four point vertex function $\Gamma_{\sigma,\sigma'}(\omega,\omega')$, which is a special case of the more general irreducible four point vertex function $\Gamma_{\sigma,\sigma',\sigma'',\sigma'''}(\omega,\omega';\omega'',\omega''')$ with $\sigma'' = \sigma$, $\sigma''' = \sigma'$, $\omega'' = \omega$ and $\omega''' = \omega'$. Our aim is to reorganise this perturbation expansion into a more convenient form to consider the strong correlation regime which corresponds to a model with $U$ large, at low temperatures, and with weak magnetic fields.

Our first step is to write the self-energy in the form,

$$\Sigma_\sigma(\omega) = \Sigma_\sigma(0) + \omega\Sigma'_\sigma(0) + \Sigma^{\text{rem}}_\sigma(\omega), \tag{31}$$

which is simply a definition for the remainder self-energy $\Sigma^{\text{rem}}_\sigma(\omega)$. Using this expression the Green's function given in equation (2) in the wide band limit can be written in the form,

$$G_{d\sigma}(\omega) = \frac{z}{\omega - \tilde{\epsilon}_d + i\tilde{\Delta} - \tilde{\Sigma}_\sigma(\omega)}. \tag{32}$$

We assume that the general theorem of Luttinger [6], that the imaginary part of $\Sigma(\omega)$ vanishes as $\omega^2$ at $\omega = 0$, so that $z$ is a real quantity. The 'renormalized' quantities, $\tilde{\epsilon}$, $\tilde{\Delta}$, are defined by (29), and the renormalized self-energy $\tilde{\Sigma}(\omega)$ by

$$\tilde{\Sigma}_\sigma(\omega) = z\Sigma^{\text{rem}}_\sigma(\omega). \tag{33}$$

The next step is to introduce rescaled creation and annihilation operators for the d-electron via

$$c^\dagger_{d,\sigma} = \sqrt{z}\tilde{c}^\dagger_{d,\sigma} \quad c_{d,\sigma} = \sqrt{z}\tilde{c}_{d,\sigma}. \tag{34}$$

We can now rewrite the Hamiltonian (1) in the form,

$$H = \tilde{H}_{\text{qp}} - \tilde{H}_{\text{c}}, \tag{35}$$

where $\tilde{H}_{\text{qp}}$ will be referred to as the quasiparticle Hamiltonian which can be written as $\tilde{H}_{\text{qp}}^{(0)} + \tilde{H}_{\text{qp}}^{(I)}$. The Hamiltonian $\tilde{H}_{\text{qp}}^{(0)}$ describes non-interacting particles and is given by

$$\tilde{H}_{\text{qp}}^{(0)} = \sum_\sigma \tilde{\epsilon}_{d,\sigma} \tilde{c}_{d,\sigma}^\dagger \tilde{c}_{d,\sigma} + \sum_{\mathbf{k},\sigma}(\tilde{V}_{\mathbf{k}} \tilde{c}_{d,\sigma}^\dagger c_{\mathbf{k},\sigma} + \tilde{V}_{\mathbf{k}}^* c_{\mathbf{k},\sigma}^\dagger \tilde{c}_{d,\sigma}) + \sum_{\mathbf{k},\sigma} \epsilon_{\mathbf{k},\sigma} c_{\mathbf{k},\sigma}^\dagger c_{\mathbf{k},\sigma}, \qquad (36)$$

and $\tilde{H}_{\text{qp}}^{(I)}$ is the interaction term,

$$\tilde{H}_{\text{qp}}^{(I)} = \tilde{U} \tilde{n}_{d,\uparrow} \tilde{n}_{d,\downarrow}, \qquad (37)$$

with $\tilde{U}$ given by (29). The second term in equation (35) will be known as the counter term and takes the form,

$$\tilde{H}_c = \lambda_1 \sum_\sigma \tilde{c}_{d,\sigma}^\dagger \tilde{c}_{d,\sigma} + \lambda_2 \tilde{n}_{d,\uparrow} \tilde{n}_{d,\downarrow}, \qquad (38)$$

where $\lambda_1$ and $\lambda_2$ are given by

$$\lambda_1 = z\Sigma(0,0), \quad \lambda_2 = z^2(\Gamma_{\uparrow,\downarrow}(0,0) - U) \qquad (39)$$

Equations (36)–(37) are simply a rewriting of the original Hamiltonian (10), so equation (35) is an identity.

We note that by construction the renormalized self-energy $\tilde{\Sigma}_\sigma(\omega)$ is such that

$$\tilde{\Sigma}_\sigma(0,0) = 0, \quad \tilde{\Sigma}_\sigma'(0,0) = 0, \qquad (40)$$

so that $\tilde{\Sigma}_\sigma(\omega) = O(\omega^2)$ for small $\omega$, on the assumption that it is analytic at $\omega = 0$. As $\tilde{\Gamma}_{\sigma,\sigma}(0,0) = 0$ we also have

$$\tilde{\Gamma}_{\sigma,\sigma'}(0,0) = \tilde{U}(1 - \delta_{\sigma,\sigma'}). \qquad (41)$$

To develop a theory appropriate for the low temperature regime we follow the renormalization procedure as used in quantum field theory so that we can make a perturbation expansion in terms of our fully dressed quasiparticles (see for instance reference [33]). We take our renormalized parameters $\tilde{\epsilon}_d$, $\tilde{\Delta}$ and $\tilde{U}$ as known and reorganise the perturbation expansion in powers of the renormalized coupling $\tilde{U}$. The full interaction Hamiltonian is $\tilde{H}_{\text{qp}}^{(I)} - \tilde{H}_c$. The terms $\lambda_1$, $\lambda_2$ and $z$ are formally expressed as series in powers of $\tilde{U}$,

$$\lambda_1 = \sum_{n=0}^\infty \lambda_1^{(n)} \tilde{U}^n, \quad \lambda_2 = \sum_{n=0}^\infty \lambda_2^{(n)} \tilde{U}^n, \quad z = \sum_{n=0}^\infty z^{(n)} \tilde{U}^n. \qquad (42)$$

The coefficients $\lambda_1^{(n)}$, $\lambda_2^{(n)}$ and $z^{(n)}$ are determined by the requirement that conditions (40) and (41) are satisfied to each order in the expansion. The perturbation expansion is about the free quasiparticle Hamiltonian given in equation (36) so that the non-interacting propagator in the expansion for the thermal Green's function is

$$\tilde{G}_{\mathrm{d}\sigma}^{(0)}(i\omega_n) = \frac{1}{i\omega_n - \tilde{\epsilon}_\mathrm{d} + i\tilde{\Delta}\mathrm{sgn}(\omega_n)}, \qquad (43)$$

where $\omega_n = (2n+1)\pi/\beta$ and $\beta = 1/k_B T$. The retarded function (30) is deduced by analytic continuation.

We propose to make a direct contact with both the Landau phenomenological and the microscopic formulations of Fermi liquid theory. The occupation number of the d-level can be deduced from the Friedel sum rule [34] which takes the form, $n_{d,\sigma}$ at $T=0$ and has the form

$$n_{\mathrm{d},\sigma}(H) = \frac{1}{2} - \frac{1}{\pi}\tan^{-1}\left(\frac{\epsilon_{\mathrm{d},\sigma} + \Sigma_\sigma(0,H)}{\Delta}\right), \qquad (44)$$

in terms of the self-energy $\Sigma(\omega)$ (in a finite magnetic field $H$) for the expansion in powers of $U$ for the 'bare' Hamiltonian (10). It takes the same form for the renormalized expansion,

$$n_{\mathrm{d},\sigma}(H) = \frac{1}{2} - \frac{1}{\pi}\tan^{-1}\left(\frac{\tilde{\epsilon}_{\mathrm{d},\sigma} + \tilde{\Sigma}_\sigma(0,H)}{\tilde{\Delta}}\right). \qquad (45)$$

This follows directly from (44) by writing it in terms of $\tilde{\epsilon}_{\mathrm{d},\sigma}$, $\tilde{\Delta}$ and $\tilde{\Sigma}$ from equation (33) as the common factor of $z$ in the argument of (45) cancels. The quasiparticle interaction plays no role as $T \to 0$ and $H \to 0$ as $\tilde{\Sigma}(0,0) = 0$ and in this limit $n_{d,\sigma}$ corresponds to the non-interacting quasiparticle number. As the effects of the quasiparticle interactions go to zero as $T \to 0$ due to the cancellation with the counter term giving $\tilde{\Sigma}(0,0) = 0$ and $\tilde{\Sigma}'(0,0) = 0$, the specific heat coefficient of the impurity $\gamma_{\mathrm{imp}}$ is due to the non-interacting quasiparticles so we obtain the free electron result (21).

Other thermodynamic results for the low temperature regime can be obtained from the lowest order term in the renormalized perturbation for $\tilde{\Sigma}$, the tadpole diagram shown in figure 6(a), which gives

$$\tilde{\Sigma}^{(1)}(\omega, H, T) = \tilde{U}(n_{d,\bar{\sigma}}^{(0)}(0, H, T) - n_{d,\sigma}^{(0)}(0, 0, 0)). \qquad (46)$$

There is no wavefunction renormalization to this order so $z^{(1)} = 0$ and also to this order $\tilde{\Gamma}^{(1)}(\omega, \omega') = \tilde{U}$, $\lambda_2^{(1)} = 0$ and $\lambda_1^{(1)} = \tilde{U}n_{d,\sigma}^{(0)}(0,0)$. The complete cancellation by the counter term only occurs for $H = 0$ and $T = 0$. The impurity spin susceptibility at $T = 0$ of the impurity to first order in $\tilde{U}$ can be calculated from $g\mu_B(n_{\mathrm{d},\uparrow} - n_{\mathrm{d},\downarrow})/2$, by substituting the self-energy from (46) into equation (45), and then differentiating with respect to $H$, to give the earlier result (22), and similarly for the charge susceptibility (23). It is not obvious that the results to first order in $\tilde{U}$ for these quantities should be exact. However, there are renormalized Ward identities [32][25] which can be derived from charge and spin conservation,

$$\left.\frac{\partial \tilde{\Sigma}_\sigma(\omega)}{\partial h}\right|_{\omega=0} = \left.\frac{\partial \tilde{\Sigma}_\sigma(\omega)}{\partial \mu}\right|_{\omega=0} = -\tilde{\rho}_{d,\sigma}(0)\tilde{U}, \quad (47)$$

where $h = g\mu_B H/2$. The spin and charge susceptibilities can be derived from the exact relations,

$$\chi_{\rm imp} = \frac{(g\mu_{\rm B})^2}{2}\tilde{\rho}_d(0)(1 - \partial\tilde{\Sigma}/\partial h) = \frac{(g\mu_{\rm B})^2}{2}\tilde{\rho}_d(0)(1 + \tilde{U}\tilde{\rho}_d(0)). \quad (48)$$

and

$$\chi_{\rm c,imp} = 2\tilde{\rho}_d(0)(1 + \partial\tilde{\Sigma}/\partial\mu) = 2\tilde{\rho}_d(0)(1 - \tilde{U}\tilde{\rho}_d(0)). \quad (49)$$

using (45), (46) and (47), confirming that the earlier results. These relations show that there are no higher order contributions in the renormalized expansion in $\tilde{U}$ for these quantities beyond first order.

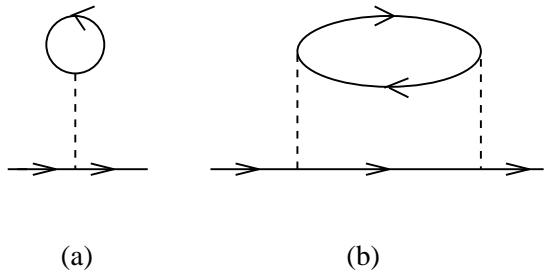

Figure 6. The two lowest order diagrams in the renormalized perturbation theory.

Exact results for the impurity Green's function (32) to order $\omega^2$ follow from the calculation of the second order diagram for $\tilde{\Sigma}$ shown in figure 6(b). There is no second order counter term contribution to this diagram from $\lambda_2$ as $\lambda_2^{(1)} = 0$. There is a contribution to $z$ which is required to eliminate the contribution from the linear term in $\omega$ to this order and this is given by

$$z^{(2)} = \frac{(\pi^2 - 12)}{4\pi^2\tilde{\Delta}^2}. \quad (50)$$

Calculation of this diagram to order $\omega^2$ gives

$$\mathrm{Im}\,\tilde{\Sigma}_\sigma(\omega, 0) = \frac{\tilde{U}^2\omega^2}{2\tilde{\Delta}(\pi\tilde{\Delta})^2} + \mathrm{O}(\omega^4). \quad (51)$$

The spectral density of the non-interacting renormalized d-Green's function describes the Kondo resonance and the $\omega^2$ terms in $\tilde{\Sigma}_\sigma(\omega, 0)$ give the low frequency corrections to this picture.

There is a temperature dependent contribution to $\tilde{\Sigma}_\sigma(0,T)$ to first order in $\tilde{U}$ given by (46). For the particle-hole symmetric model, and more generally in the Kondo regime $n_d \to 1$, this vanishes as $n_{d,\sigma}(0,T) = n_{d,\sigma}(0,0) = 1$ and the leading order temperature dependence arises from the second order diagram 1(b). This gives the result,

$$\mathrm{Im}\,\tilde{\Sigma}_\sigma(0,T) = \frac{(\pi\tilde{U})^2 k_B^2 T^2}{2\tilde{\Delta}(\pi\tilde{\Delta})^2} + \mathrm{O}(T^4). \tag{52}$$

These results can be used to calculate the leading order $T^2$ contribution to the impurity conductivity $\sigma_{\mathrm{imp}}(T)$ which is given by

$$\sigma_{\mathrm{imp}}(T) = \sigma_0 \left\{ 1 + \frac{\pi^2}{3}\left(\frac{k_\mathrm{B} T}{\tilde{\Delta}}\right)^2 (1 + 2(R-1)^2) + \mathrm{O}(T^4) \right\}. \tag{53}$$

where $R$ is the Wilson ratio given by

$$R = 1 + \tilde{U}/\pi\tilde{\Delta}. \tag{54}$$

This conductivity result was first derived by Nozières for the Kondo regime [30] and for the more general case by Yamada [25]. The formula for the Wilson ratio is a more general one than that given earlier and goes over to (26) $R \to 2$ in the Kondo regime on using the relation (25) between the renormalized parameters. Hence all the basic Fermi liquid results can be obtained within the renormalized expansion up to second order in $\tilde{U}$. The wavefunction renormalization factor $z$ is required to relate the bare spectral density $\rho_d(\omega)$ to the quasiparticle density of states $\tilde{\rho}_d(\omega)$ but does not enter explicitly the calculation of the thermodynamics which can be expressed entirely in terms of the renormalized parameters $\tilde{\epsilon}_d$, $\tilde{\Delta}$ and $\tilde{U}$.

We see that the effects of the counter terms play no really significant role in the Fermi liquid regime. The renormalized perturbation theory with the counter terms, however, goes beyond the Fermi liquid regime. In this perturbation theory nothing has been omitted so that in principle calculations can be performed at high temperatures and high fields allowing the bare particles to be seen. If the primary interest is in these regimes then the more appropriate starting point is the model in terms of the bare parameters rather than the renormalized ones. It should be possible to estimate the renormalized parameters from the bare ones by a separate variational calculation for the single and two particle excitations from the ground state. In principle it is possible from the results of the renormalized perturbation theory to estimate the bare parameters from the renormalized ones by inverting (39) and (42),

$$\epsilon_d = (\tilde{\epsilon}_d - \lambda_1)/z, \quad \Delta = \tilde{\Delta}/z, \quad U = (\tilde{U} - \lambda_2)/z^2, \tag{55}$$

where $\lambda_1$, $\lambda_2$ and $z$ are implicit functions of $\tilde{\epsilon}_d$, $\tilde{\Delta}$ and $\tilde{U}$. In the very weak coupling limit $\tilde{U} \ll \pi\tilde{\Delta}$ $z \to 1$ and $\epsilon_d \to \tilde{\epsilon}_d$, $\Delta \to \tilde{\Delta}$ and $U \to \tilde{U}$, as we can seen in the results for the symmetric model in figure 2. The most useful regime for the renormalized approach, however, should be in calculating corrections to Fermi liquid theory. The calculation with counter terms gives a systematic procedure for taking such corrections into account.

# 4 Renormalized Perturbation Theory for Translationally Invariant Systems

The renormalized perturbation approach described in the previous section gives us a very clear picture of the physics of the low temperature regime for a magnetic impurity as described by the Anderson model. It combines elements of the renormalization group, microscopic and phenomenological Fermi liquid approaches, as well retaining many of the more intuitive elements of the original Landau theory. In this section we investigate whether it is possible to generalize the approach to a system which has translational invariance, and deal with the complications which arise when the self-energy depends on the momentum vector $\mathbf{k}$ as well as the frequency $\omega$. We consider a quite general model of fermions in states classified by momenta $\mathbf{k}$ and spin $\sigma$ described by a Hamiltonian,

$$H = \sum_{\mathbf{k},\sigma} \epsilon_{\mathbf{k},\sigma} c^\dagger_{\mathbf{k},\sigma} c_{\mathbf{k},\sigma} + \frac{1}{2} \sum_{\mathbf{k_1},\mathbf{k_2},\sigma_1,\sigma_2} V(\mathbf{q}) c^\dagger_{\mathbf{k_1}+\mathbf{q},\sigma_1} c^\dagger_{\mathbf{k_2}-\mathbf{q},\sigma_2} c_{\mathbf{k_2},\sigma_2} c_{\mathbf{k_1},\sigma_1}, \qquad (56)$$

where $\epsilon_{\mathbf{k}} = \mathbf{k}^2/2m$ with $m$ the mass, and an interaction $V(\mathbf{q})$ which is a function of the momentum transfer $\mathbf{q}$. It will be useful to express the Hamiltonian in a more general form,

$$H = \sum_{\mathbf{k},\sigma} \epsilon_{\mathbf{k},\sigma} c^\dagger_{\mathbf{k},\sigma} c_{\mathbf{k},\sigma} + \frac{1}{4} \sum_{\mathbf{k}s,\sigma s} \Gamma^{(0)}_{\sigma_1\sigma_2\sigma_3\sigma_4}(\mathbf{k_1},\mathbf{k_2},\mathbf{k_3},\mathbf{k_4}) c^\dagger_{\mathbf{k_3},\sigma_3} c^\dagger_{\mathbf{k_4},\sigma_4} c_{\mathbf{k_2},\sigma_2} c_{\mathbf{k_1},\sigma_1}, \qquad (57)$$

where the interaction is antisymmetric under the exchanges $(\mathbf{k}_1,\sigma_1) \leftrightarrow (\mathbf{k}_2,\sigma_2)$ and $(\mathbf{k}_3,\sigma_3) \leftrightarrow (\mathbf{k}_4,\sigma_4)$. For translational invariance and spin conservation we have

$$\mathbf{k_1} + \mathbf{k_2} = \mathbf{k_3} + \mathbf{k_4} \quad \sigma_1 + \sigma_2 = \sigma_3 + \sigma_4. \qquad (58)$$

For a spin independent interaction depending only on the momentum transfer $\mathbf{q}$ as in(56) with $V(\mathbf{q}) = V(-\mathbf{q})$, then the antisymmetrized form is

$$\Gamma^{(0)}_{\sigma_1\sigma_2\sigma_3\sigma_4}(\mathbf{k}_1,\mathbf{k}_2,\mathbf{k}_3,\mathbf{k}_4) = \delta_{\sigma_1,\sigma_3}\delta_{\sigma_2,\sigma_4} V(\mathbf{k}_3-\mathbf{k}_1) - \delta_{\sigma_1,\sigma_4}\delta_{\sigma_2,\sigma_3} V(\mathbf{k}_4-\mathbf{k}_1). \qquad (59)$$

We can then write (56) as

$$H = \sum_{\mathbf{k},\sigma} \epsilon_{\mathbf{k},\sigma} c^\dagger_{\mathbf{k},\sigma} c_{\mathbf{k},\sigma} +$$
$$\frac{1}{4} \sum_{\mathbf{k_1},\mathbf{k_2},\mathbf{q},\sigma s} \Gamma^{(0)}_{\sigma_1\sigma_2\sigma_3\sigma_4}(\mathbf{k}_1,\mathbf{k}_2,\mathbf{k}_1+\mathbf{q},\mathbf{k}_2-\mathbf{q}) c^\dagger_{\mathbf{k_1}+\mathbf{q},\sigma_3} c^\dagger_{\mathbf{k_2}-\mathbf{q},\sigma_4} c_{\mathbf{k_2},\sigma_2} c_{\mathbf{k_1},\sigma_1}, \qquad (60)$$

The interacting one electron Green's function can be expressed in the usual form,

$$G_\sigma(\mathbf{k},\omega) = \frac{1}{\omega - \epsilon_{\mathbf{k}} - \Sigma(\mathbf{k},\omega)}, \qquad (61)$$

where $\Sigma(\mathbf{k},\omega)$ is the corresponding self-energy. To follow our previous prescription we expand the self-energy in a Taylor's series to first order about the Fermi surface, and keep the remainder term,

$$\Sigma(k,\omega) = \Sigma(k_F, \epsilon_F) + (\omega - \epsilon_F)\frac{\partial \Sigma(k_F, \epsilon_F)}{\partial \omega} + (k - k_F)\frac{\partial \Sigma(k_F, \epsilon_F)}{\partial k} + \Sigma^{\rm rem}(\mathbf{k},\omega) \quad (62)$$

where we have assumed $\epsilon_k = \epsilon(|k|)$. We can rewrite this as

$$G_\sigma(k,\omega) = \frac{z}{\omega - \tilde{\epsilon}_k - \tilde{\Sigma}(k,\omega)} \quad (63)$$

with $z$, the wavefunction renormalization factor on the Fermi surface, given by

$$z = \frac{1}{1 - \frac{\partial \Sigma(k_F, \epsilon_F)}{\partial \omega}}, \quad (64)$$

and with the renormalized energies given by

$$\tilde{\epsilon}_k = z\left((\epsilon_k - \epsilon_F) + (k - k_F)\frac{\partial \Sigma(k_F, \epsilon_F)}{\partial k} + \Sigma(k_F, \epsilon_F)\right). \quad (65)$$

Similarly to (33) we define a renormalized self-energy by

$$\tilde{\Sigma}(k,\omega) = z\Sigma^{\rm rem}(k,\omega), \quad (66)$$

with $\omega$ measured relative to the Fermi level. We introduce rescaled operators,

$$c^\dagger_{\mathbf{k},\sigma} = \sqrt{z}\tilde{c}^\dagger_{\mathbf{k},\sigma}, \quad c_{\mathbf{k},\sigma} = \sqrt{z}\tilde{c}_{\mathbf{k},\sigma}, \quad (67)$$

and define a quasiparticle Green's function,

$$\tilde{G}_\sigma(k,\omega) = \frac{1}{\omega - \tilde{\epsilon}_\mathbf{k} - \tilde{\Sigma}(k,\omega)}. \quad (68)$$

The full irreducible renormalized four vertex is defined by

$$\tilde{\Gamma}_{\sigma_1\sigma_2\sigma_3\sigma_4}(\mathbf{k}_1,\omega_1,\mathbf{k}_2,\omega_2,\mathbf{k}_1+\mathbf{q},\omega_1+\omega,\mathbf{k}_2-\mathbf{q},\omega_2-\omega) =$$
$$z^2 \Gamma_{\sigma_1\sigma_2\sigma_3\sigma_4}(\mathbf{k}_1,\omega_1,\mathbf{k}_2,\omega_2,\mathbf{k}_1+\mathbf{q},\omega_1+\omega,\mathbf{k}_2-\mathbf{q},\omega_2-\omega). \quad (69)$$

So far every step has been an obvious generalization of the impurity case. When it comes to defining the renormalized interaction we have to be a little careful. Due to repeated particle–hole scattering the vertex has a singularity such that there is no unique way of taking the limits $\omega \to 0$ and $\mathbf{q} \to 0$. We leave this problem for the moment and assume that we can find a suitable antisymmetrized renormalized quasiparticle interaction $\tilde{V}^{\sigma_1\sigma_2}_{\sigma_3\sigma_4}(\mathbf{k}_1,\mathbf{k}_2,\mathbf{k}_1+\mathbf{q},\mathbf{k}_2-\mathbf{q})$ and hence construct a quasiparticle Hamiltonian,

$$\tilde{H}_{\rm qp} = \sum_{\mathbf{k},\sigma} \tilde{\epsilon}_{\mathbf{k},\sigma} \tilde{c}^\dagger_{\mathbf{k},\sigma} \tilde{c}_{\mathbf{k},\sigma} +$$
$$\frac{1}{4} \sum_{\mathbf{k}_1,\mathbf{k}_2,\mathbf{q},\sigma s} \tilde{V}^{\sigma_1\sigma_2}_{\sigma_3\sigma_4}(\mathbf{k}_1,\mathbf{k}_2,\mathbf{k}_1+\mathbf{q},\mathbf{k}_2-\mathbf{q}) \tilde{c}^\dagger_{\mathbf{k}_1+\mathbf{q},\sigma_3} \tilde{c}^\dagger_{\mathbf{k}_2-\mathbf{q},\sigma_4} \tilde{c}_{\mathbf{k}_2,\sigma_2} \tilde{c}_{\mathbf{k}_1,\sigma_1}. \quad (70)$$

As earlier, the ground state is defined as the vacuum state with no excitations and the Hamiltonian has to be normal ordered and expressed in terms of quasiparticles and quasiholes so that the interaction only comes into play when more than one excitation is created from the (interacting) ground state. Following the steps we used in the previous section we write our original Hamiltonian (56) in the form,

$$H = \tilde{H}_{\mathrm{qp}} - \tilde{H}_{\mathrm{c}}, \tag{71}$$

where $\tilde{H}_{\mathrm{c}}$ is the Hamiltonian for the counter terms and takes the form,

$$\tilde{H}_{\mathrm{c}} = \sum_{\mathbf{k},\sigma}(\lambda^{(a)} + \lambda^{(b)}(k - k_{\mathrm{F}}))\tilde{c}^\dagger_{\mathbf{k},\sigma}\tilde{c}_{\mathbf{k},\sigma} +$$
$$\frac{1}{4}\sum_{\mathbf{k}_1,\mathbf{k}_2,\mathbf{q},\sigma s}\lambda^{(c)}_{\sigma_1\sigma_2\sigma_3\sigma_4}(\mathbf{k}_1,\mathbf{k}_2,\mathbf{k}_1+\mathbf{q},\mathbf{k}_2-\mathbf{q})\tilde{c}^\dagger_{\mathbf{k}_1+\mathbf{q},\sigma_3}\tilde{c}^\dagger_{\mathbf{k}_2-\mathbf{q},\sigma_4}\tilde{c}_{\mathbf{k}_2,\sigma_2}\tilde{c}_{\mathbf{k}_1,\sigma_1}. \tag{72}$$

where

$$\lambda^{(a)} = z\frac{\partial \Sigma(k_{\mathrm{F}},\epsilon_{\mathrm{F}})}{\partial k}, \quad \lambda^{(b)} = z(\Sigma(k_{\mathrm{F}},\epsilon_{\mathrm{F}}) - \epsilon_{\mathrm{F}}) \tag{73}$$

and

$$\lambda^{(c)}_{\sigma_1\sigma_2\sigma_3\sigma_4}(\mathbf{k}_1,\mathbf{k}_2,\mathbf{k}_1+\mathbf{q},\mathbf{k}_2-\mathbf{q}) =$$
$$\tilde{V}^{\sigma_1\sigma_2}_{\sigma_3\sigma_4}(\mathbf{k}_1,\mathbf{k}_2,\mathbf{k}_1+\mathbf{q},\mathbf{k}_2-\mathbf{q}) - z^2\Gamma^{(0)}_{\sigma_1\sigma_2\sigma_3\sigma_4}(\mathbf{k}_1,\mathbf{k}_2,\mathbf{k}_1+\mathbf{q},\mathbf{k}_2-\mathbf{q}) \tag{74}$$

We can define a renormalized perturbation expansion, as earlier, in powers of $\tilde{V}$ with the coefficients of the counter terms and the $z$ factor chosen to satisfy the conditions,

$$\tilde{\Sigma}(k_{\mathrm{F}},\epsilon_{\mathrm{F}}) = 0, \quad \frac{\partial \tilde{\Sigma}(k_{\mathrm{F}},\epsilon_{\mathrm{F}})}{\partial k} = 0, \quad \tilde{\Sigma}'(k_{\mathrm{F}},\epsilon_{\mathrm{F}}) = 0, \tag{75}$$

for zero field and at zero temperature. These conditions reflect the fact that the quasiparticle energies and Fermi level used in the quasiparticle Hamiltonian are already fully renormalized. We now consider how to choose $\tilde{V}^{\sigma_1\sigma_2}_{\sigma_3\sigma_4}(\mathbf{k}_1,\mathbf{k}_2,\mathbf{k}_1+\mathbf{q},\mathbf{k}_2-\mathbf{q})$ and how to fix the counter term $\lambda^{(c)}_{\sigma_1\sigma_2\sigma_3\sigma_4}(\mathbf{k}_1,\mathbf{k}_2,\mathbf{k}_1+\mathbf{q},\mathbf{k}_2-\mathbf{q})$. In the renormalized perturbation approach the renormalized interaction $\tilde{V}^{\sigma_1\sigma_2}_{\sigma_3\sigma_4}(\mathbf{k}_1,\mathbf{k}_2,\mathbf{k}_1+\mathbf{q},\mathbf{k}_2-\mathbf{q})$ is a matter of choice, because it is added and subtracted, but there should be an optimum choice for calculations of the low temperature behaviour as there was in the impurity case. An obvious choice is

$$\tilde{V}^{\sigma_1\sigma_2}_{\sigma_3\sigma_4}(\mathbf{k}_1,\mathbf{k}_2,\mathbf{k}_1+\mathbf{q},\mathbf{k}_2-\mathbf{q}) = \tilde{\Gamma}_{\sigma_1\sigma_2\sigma_3\sigma_4}(\mathbf{k}_1,0,\mathbf{k}_2,0,\mathbf{k}_1+\mathbf{q},0,\mathbf{k}_2-\mathbf{q},0) \tag{76}$$

which is the quasiparticle scattering function. Its antisymmetry follows from its definition. When we come to evaluate the lowest order Hartree diagram of the renormalized theory, this has to be evaluated in the zero momentum exchange limit $\mathbf{q} \to 0$. For a more general retarded $\omega$ dependent interaction this diagram has to be evaluated in the limit $\mathbf{q} \to 0$ followed by $\omega \to 0$ (there is a small imaginary contribution to $\omega$ associated with the time ordering for this diagram). Repeated quasiparticle-quasihole scattering gives a singular contribution to the retarded interaction and it is important to take this into account before we take the limits so that we can take them in the correct order. Similarly the Fock term, which is evaluated for $\mathbf{q} \to \mathbf{k}_2 - \mathbf{k}_1$ followed by $\omega \to 0$, also has a singular contribution from quasiparticle-quasihole scattering.

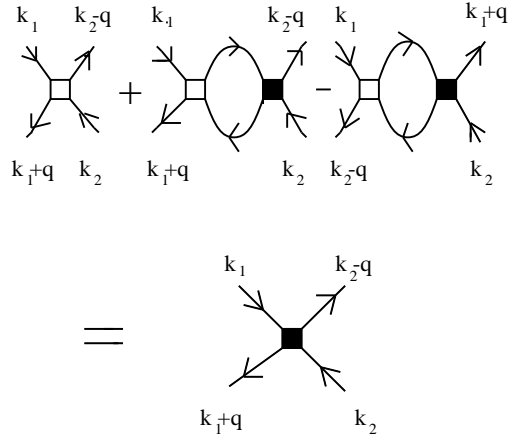

Figure 7. Particle-hole diagrams corresponding to the integral equation (77). The unfilled square vertex corresponds to $U^{\sigma_1 \sigma_2}_{\sigma_3 \sigma_4}(\mathbf{k}_1, \mathbf{k}_2, \mathbf{k}_1 + \mathbf{q}, \mathbf{k}_2 - \mathbf{q})$ and the square filled vertex to $\tilde{\Gamma}^{\text{p-h}}_{\sigma_1 \sigma_2 \sigma_3 \sigma_4}(\mathbf{k}_1, \mathbf{k}_2, \mathbf{k}_1 + \mathbf{q}, \mathbf{k}_2 - \mathbf{q}, \omega)$.

If we denote the renormalized vertex corresponding to the sum the quasiparticle-quasihole series shown in figure 7 by $\tilde{\Gamma}^{\text{p-h}}_{\sigma_1 \sigma_2 \sigma_3 \sigma_4}(\mathbf{k}_1, \mathbf{k}_2, \mathbf{k}_1 + \mathbf{q}, \mathbf{k}_2 - \mathbf{q}, \omega)$ then we obtain the integral equation,

$$\tilde{\Gamma}^{\text{p-h}}_{\sigma_1 \sigma_2 \sigma_3 \sigma_4}(\mathbf{k}_1, \mathbf{k}_2, \mathbf{k}_1 + \mathbf{q}, \mathbf{k}_2 - \mathbf{q}, \omega) = U^{\sigma_1 \sigma_2}_{\sigma_3 \sigma_4}(\mathbf{k}_1, \mathbf{k}_2, \mathbf{k}_1 + \mathbf{q}, \mathbf{k}_2 - \mathbf{q})$$
$$+ \sum_{\sigma' \sigma''} \int U^{\sigma_1 \sigma''}_{\sigma_3 \sigma'}(\mathbf{k}_1, \mathbf{k}'', \mathbf{k}_1 + \mathbf{q}, \mathbf{k}'' - \mathbf{q}) F(\mathbf{k}'', \mathbf{q}; \omega) \tilde{\Gamma}^{\text{p-h}}_{\sigma' \sigma_2 \sigma'' \sigma_4}(\mathbf{k}'' - \mathbf{q}, \mathbf{k}_2, \mathbf{k}'', \mathbf{k}_2 - \mathbf{q}, \omega) d\mathbf{k}''$$
$$- \sum_{\sigma' \sigma''} \int U^{\sigma_1 \sigma''}_{\sigma_4 \sigma'}(\mathbf{k}_1, \mathbf{k}'', \mathbf{k}_2 + \mathbf{q}, \mathbf{k}'' - \mathbf{k}_2 + \mathbf{k}_1 - \mathbf{q}) F(\mathbf{k}'', \mathbf{k}_2 - \mathbf{k}_1 - \mathbf{q}; \omega)$$
$$\tilde{\Gamma}^{\text{p-h}}_{\sigma'' \sigma_2 \sigma_3 \sigma'}(\mathbf{k}'' - \mathbf{k}_2 + \mathbf{k}_1 + \mathbf{q}, \mathbf{k}_2, \mathbf{k}_1 + \mathbf{q}, \mathbf{k}'', \omega) d\mathbf{k}'' \quad (77)$$

where

$$U^{\sigma_1 \sigma_2}_{\sigma_3 \sigma_4}(\mathbf{k}_1, \mathbf{k}_2, \mathbf{k}_1 + \mathbf{q}, \mathbf{k}_2 - \mathbf{q}) =$$
$$\tilde{V}^{\sigma_1 \sigma_2}_{\sigma_3 \sigma_4}(\mathbf{k}_1, \mathbf{k}_2, \mathbf{k}_1 + \mathbf{q}, \mathbf{k}_2 - \mathbf{q}) - \lambda^{(c)}_{\sigma_1 \sigma_2 \sigma_3 \sigma_4}(\mathbf{k}_1, \mathbf{k}_2, \mathbf{k}_1 + \mathbf{q}, \mathbf{k}_2 - \mathbf{q})$$
(78)

and the propagator for the quasiparticle-quasihole pair $F(\mathbf{k}'', \mathbf{q}; \omega)$ is given by

$$F(\mathbf{k}'', \mathbf{q}; \omega) = \frac{1}{V_0 (2\pi)^3} \frac{f(\tilde{\epsilon}_{\mathbf{k}''}) - f(\tilde{\epsilon}_{\mathbf{k}'' - \mathbf{q}})}{i\omega + \tilde{\epsilon}_{\mathbf{k}''} - \tilde{\epsilon}_{\mathbf{k}'' - \mathbf{q}}}, \quad (79)$$

where $V_0$ is the volume. If we take the limit $\omega \to 0$ in (77), then as (79) does not vanish in this limit it is clear that the counter terms must be non-zero as we have implicitly already taken some of these contributions to $\tilde{\Gamma}_{\sigma_1\sigma_2\sigma_3\sigma_4}(\mathbf{k}_1, \mathbf{k}_2, \mathbf{k}_1+\mathbf{q}, \mathbf{k}_2-\mathbf{q})$ into account. We have to choose the counter term so that at $T=0$ we satisfy the condition expressed in equation (76). This implies

$$\tilde{V}^{\sigma_1\sigma_2}_{\sigma_3\sigma_4}(\mathbf{k}_1, \mathbf{k}_2, \mathbf{k}_1+\mathbf{q}, \mathbf{k}_2-\mathbf{q}) = U^{\sigma_1\sigma_2}_{\sigma_3\sigma_4}(\mathbf{k}_1, \mathbf{k}_2, \mathbf{k}_1+\mathbf{q}, \mathbf{k}_2-\mathbf{q})$$

$$+ \sum_{\sigma'\sigma''} \int U^{\sigma_1\sigma''}_{\sigma_3\sigma'}(\mathbf{k}_1, \mathbf{k}'', \mathbf{k}_1+\mathbf{q}, \mathbf{k}''-\mathbf{q}) F(\mathbf{k}'', \mathbf{q}; 0) \tilde{V}^{\sigma'\sigma_2}_{\sigma''\sigma_4}(\mathbf{k}''-\mathbf{q}, \mathbf{k}_2, \mathbf{k}'', \mathbf{k}_2-\mathbf{q}) d\mathbf{k}''$$

$$- \sum_{\sigma'\sigma''} \int U^{\sigma_1\sigma''}_{\sigma_4\sigma'}(\mathbf{k}_1, \mathbf{k}'', \mathbf{k}_2-\mathbf{q}, \mathbf{k}''-\mathbf{k}_2+\mathbf{k}_1+\mathbf{q}) F(\mathbf{k}'', \mathbf{k}_2-\mathbf{k}_1-\mathbf{q}; 0)$$

$$\tilde{V}^{\sigma''\sigma_2}_{\sigma_3\sigma'}(\mathbf{k}''+\mathbf{k}_1-\mathbf{k}_2+\mathbf{q}, \mathbf{k}_2, \mathbf{k}_1+\mathbf{q}, \mathbf{k}'') d\mathbf{k}'' \qquad (80)$$

For a given $\tilde{V}^{\sigma_1\sigma_2}_{\sigma_3\sigma_4}(\mathbf{k}_1, \mathbf{k}_2, \mathbf{k}_1+\mathbf{q}, \mathbf{k}_2-\mathbf{q})$ let us denote the solution of (80) for $U^{\sigma_1\sigma_2}_{\sigma_3\sigma_4}(\mathbf{k}_1, \mathbf{k}_2, \mathbf{k}_1+\mathbf{q}, \mathbf{k}_2-\mathbf{q})$ at $T=0$ by $\tilde{U}^{\sigma_1\sigma_2}_{\sigma_3\sigma_4}(\mathbf{k}_1, \mathbf{k}_2, \mathbf{k}_1+\mathbf{q}, \mathbf{k}_2-\mathbf{q})$. The antisymmetry of $\tilde{U}^{\sigma_1\sigma_2}_{\sigma_3\sigma_4}(\mathbf{k}_1, \mathbf{k}_2, \mathbf{k}_1+\mathbf{q}, \mathbf{k}_2-\mathbf{q})$ with respect to the transformation $(\mathbf{k}_1+\mathbf{q}, \sigma_3) \leftrightarrow (\mathbf{k}_2-\mathbf{q}, \sigma_4)$ (or equivalently $(\mathbf{k}_1, \sigma_1) \leftrightarrow (\mathbf{k}_2, \sigma_2)$) follows from equation (77).

We can now show that in the limit $\mathbf{q} \to 0$ $\tilde{U}^{\sigma_1\sigma_2}_{\sigma_3\sigma_4}(\mathbf{k}_1, \mathbf{k}_2, \mathbf{k}_1+\mathbf{q}, \mathbf{k}_2-\mathbf{q})$ can be identified as the Landau quasiparticle interaction function $f^{\sigma_1\sigma_2}_{\sigma_3\sigma_4}(\mathbf{k}_1, \mathbf{k}_2)$, which is the more general form than that introduced in (1) which takes account of the possibility of spin dependent interactions.

The contribution to the direct term in the integral equation (77) has a pole arising from $F(\mathbf{k}'', \mathbf{q}; \omega)$ for small $\mathbf{q}$ and $\omega$. At $T=0$ and as $\mathbf{q} \to 0$ we have, on continuing to real frequencies ($i\omega \to \omega$),

$$F(\mathbf{k}'', \mathbf{q}; \omega) = \delta(|\mathbf{k}''| - k_{\rm F}) \frac{\mathbf{n}'' \cdot \mathbf{q}\, v_{\rm F}}{(\omega - \mathbf{n}'' \cdot \mathbf{q}\, v_{\rm F})} \qquad (81)$$

where $v_{\rm F} = k_{\rm F}/m$ and $\mathbf{n}'' = -\mathbf{k}''/|\mathbf{k}''|$.

We can follow the standard treatment given in reference [8], chapter 4, and represent $\tilde{\Gamma}^{\rm p-h}_{\sigma_1\sigma_2\sigma_3\sigma_4}(\mathbf{k}_1, \mathbf{k}_2, \mathbf{k}_1+\mathbf{q}, \mathbf{k}_2-\mathbf{q}, \omega)$ by $\chi_{\sigma_1\sigma_3}(\mathbf{k}_1, \mathbf{q}, \omega) \chi'_{\sigma_2\sigma_4}(\mathbf{k}_2, \mathbf{q}, \omega)$ near the pole. If we keep only the dominant terms for small $\mathbf{q}$ and $\omega$ we can cancel the common factor $\chi'_{\sigma_2\sigma_4}(\mathbf{k}_2, \mathbf{q}, \omega)$ and then write

$$\nu_{\sigma_1\sigma_3}(\mathbf{n}_1) = \frac{\mathbf{n}_1 \cdot \mathbf{q}}{(\omega - \mathbf{n}_1 \cdot \mathbf{q}\, v_{\rm F})} \chi_{\sigma_1\sigma_3}(\mathbf{k}_1, \mathbf{q}, \omega), \qquad (82)$$

where $\mathbf{n}_1$ is a unit vector in the direction of $\mathbf{k}_1$, so that for $\mathbf{q} \to 0$ we find

$$(\omega - \mathbf{n}_1 \cdot \mathbf{q}\, v_{\rm F}) \nu_{\sigma_1\sigma_3}(\mathbf{n}_1) = \mathbf{n}_1 \cdot \mathbf{q} \frac{k_{\rm F}^2}{(2\pi)^3} \int \tilde{U}^{\sigma_1\sigma_2}_{\sigma_3\sigma_4}(\mathbf{k}_1, \mathbf{k}'', \mathbf{k}_1, \mathbf{k}'') \nu_{\sigma_2\sigma_4}(\mathbf{n}'') d\Omega'', \qquad (83)$$

where the integration is over the solid angle $\Omega''$. Equation (83) corresponds to the Landau equation for the zero sound collective modes (see[8]) so it enables us to make the identification,

$$f^{\sigma_1\sigma_2}_{\sigma_3\sigma_4}(\mathbf{k}_1, \mathbf{k}_2) = \tilde{U}^{\sigma_1\sigma_2}_{\sigma_3\sigma_4}(\mathbf{k}_1, \mathbf{k}_2, \mathbf{k}_1, \mathbf{k}_2). \qquad (84)$$

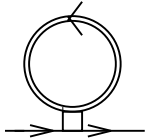

Figure 8. The Hartree-Fock diagram in the renormalized perturbation theory. The vertex corresponds to $\tilde{U}^{\sigma_1\sigma_2}_{\sigma_3\sigma_4}(\mathbf{k}_1, \mathbf{k}_2, \mathbf{k}_1+\mathbf{q}, \mathbf{k}_2-\mathbf{q})$ and the double line represents the dressed propagator. There is also a counter term arising from the first term in (72).

In the renormalized perturbation expansion we should include the more general vertex with the repeated quasiparticle-quasihole scattering given by (77) in evaluating the Hartree (-Fock) diagram, figure 8, so that we take account of all the singular contributions in the $\omega \to 0$ limit. We then get the correct results when we take the limits $\mathbf{q} \to 0$ ($\mathbf{q} \to \mathbf{k}_2 - \mathbf{k}_1$) followed by $\omega \to 0$. Evaluation of the Hartree-Fock diagram is equivalent to taking the mean field approximation on the quasiparticle Hamiltonian,

$$\tilde{H}_{\text{qp}} = \sum_{\mathbf{k},\sigma} \tilde{\epsilon}_{\mathbf{k},\sigma} \tilde{c}^\dagger_{\mathbf{k},\sigma} \tilde{c}_{\mathbf{k},\sigma} +$$
$$\frac{1}{4} \sum_{\mathbf{k}_1,\mathbf{k}_2,\mathbf{q},\sigma s} \tilde{U}^{\sigma_1\sigma_2}_{\sigma_3\sigma_4}(\mathbf{k}_1, \mathbf{k}_2, \mathbf{k}_1+\mathbf{q}, \mathbf{k}_2-\mathbf{q}) \tilde{c}^\dagger_{\mathbf{k}_1+\mathbf{q},\sigma_3} \tilde{c}^\dagger_{\mathbf{k}_2-\mathbf{q},\sigma_4} \tilde{c}_{\mathbf{k}_2,\sigma_2} \tilde{c}_{\mathbf{k}_1,\sigma_1}.$$
(85)

This gives the total excitation energy in the form,

$$E_{\text{tot}} = E_{\text{gs}} + \sum_{\mathbf{k},\sigma} \tilde{\epsilon}_{\mathbf{k},\sigma} \delta\tilde{n}_{\mathbf{k},\sigma} + \frac{1}{2} \sum_{\mathbf{k}_1,\mathbf{k}_2,\sigma_1,\sigma_2} \tilde{U}^{\sigma_1\sigma_2}_{\sigma_1\sigma_2}(\mathbf{k}_1,\mathbf{k}_2,\mathbf{k}_1,\mathbf{k}_2) \delta\tilde{n}_{\mathbf{k}_1,\sigma_1} \delta\tilde{n}_{\mathbf{k}_2,\sigma_2}, \quad (86)$$

when we have spin independent interactions. This equation can be identified with the Landau equation (1) as $f^{\sigma_1\sigma_2}_{\sigma_3\sigma_4}(\mathbf{k}_1,\mathbf{k}_2) = \tilde{U}^{\sigma_1\sigma_2}_{\sigma_3\sigma_4}(\mathbf{k}_1,\mathbf{k}_2,\mathbf{k}_1,\mathbf{k}_2)$.

We can choose to work with the quasiparticle Hamiltonian (85) with the interaction $\tilde{U}^{\sigma_1\sigma_2}_{\sigma_3\sigma_4}(\mathbf{k}_1,\mathbf{k}_2,\mathbf{k}_1+\mathbf{q},\mathbf{k}_2-\mathbf{q})$. This has the advantage that the mean field theory corresponds to the Landau total energy functional, as in the impurity case, and it has no explicit interaction counter term when the low energy collective quasiparticle-quasihole excitations are taken into account. It satisfies the conditions for the optimum choice of the quasiparticle Hamiltonian.

The two scattering functions, $\tilde{V}^{\sigma_1\sigma_2}_{\sigma_3\sigma_4}(\mathbf{k}_1,\mathbf{k}_2,\mathbf{k}_1+\mathbf{q},\mathbf{k}_2-\mathbf{q})$ and $\tilde{U}^{\sigma_1\sigma_2}_{\sigma_3\sigma_4}(\mathbf{k}_1,\mathbf{k}_2,\mathbf{k}_1+\mathbf{q},\mathbf{k}_2-\mathbf{q})$, are related by the integral equation (80) for $T = 0$ with $U^{\sigma_1\sigma_2}_{\sigma_3\sigma_4}(\mathbf{k}_1,\mathbf{k}_2,\mathbf{k}_1+\mathbf{q},\mathbf{k}_2-\mathbf{q}) \to \tilde{U}^{\sigma_1\sigma_2}_{\sigma_3\sigma_4}(\mathbf{k}_1,\mathbf{k}_2,\mathbf{k}_1+\mathbf{q},\mathbf{k}_2-\mathbf{q})$. It follows from the definition (76) that in the $\mathbf{q} \to 0$ limit $\tilde{V}^{\sigma_1\sigma_2}_{\sigma_3\sigma_4}(\mathbf{k}_1,\mathbf{k}_2+\mathbf{q},\mathbf{k}_1,\mathbf{k}_2-\mathbf{q})$ is the forward scattering function for two quasiparticles in the Landau theory which we denote by $A^{\sigma_1\sigma_2}_{\sigma_3\sigma_4}(\mathbf{k}_1,\mathbf{k}_2)$,

$$A^{\sigma_1\sigma_2}_{\sigma_3\sigma_4}(\mathbf{k}_1,\mathbf{k}_2) = \tilde{V}^{\sigma_1\sigma_2}_{\sigma_3\sigma_4}(\mathbf{k}_1,\mathbf{k}_2,\mathbf{k}_1,\mathbf{k}_2). \quad (87)$$

If we take the limit $\mathbf{q} \to 0$ in equation (80), however, the equation does not reduce to the usual one (see [8]) relating $A^{\sigma_1\sigma_2}_{\sigma_3\sigma_4}(\mathbf{k}_1,\mathbf{k}_2)$ and $f^{\sigma_1\sigma_2}_{\sigma_3\sigma_4}(\mathbf{k}_1,\mathbf{k}_2)$, due to the presence of the exchange term. In this respect this derivation differs from the standard derivations of the microscopic Landau theory. However, equation (80) would appear to be more satisfactory because it is applicable for general $\mathbf{q}$ and includes the singular contribution at $\mathbf{q} = \mathbf{k}_2 - \mathbf{k}_1$ arising form the pole in the exchange term in (77) at $\omega = \mathbf{n}'' \cdot (\mathbf{k}_2 - \mathbf{k}_1 - \mathbf{q})v_{\mathrm{F}}$.

To summarize the situation so far:

(i). We identify the quasiparticle Hamiltonian with the interaction $\tilde{U}^{\sigma_1\sigma_2}_{\sigma_3\sigma_4}(\mathbf{k}_1, \mathbf{k}_2, \mathbf{k}_1 + \mathbf{q}, \mathbf{k}_2 - \mathbf{q})$ as the Hamiltonian which determines the low energy excitations of a Fermi liquid. We conclude that the mean field approximation to this model corresponds to the Landau total energy functional. This Hamiltonian has non-forward scattering terms ($\mathbf{q} \neq 0$), which do not enter the Landau theory, but these do not contribute at the mean field level. For low energy quasiparticle-quasihole pairs in the region of the Fermi surface there is only a rather restricted phase space for such scattering. In three dimensions with a spherical Fermi surface and states $\mathbf{k}_1, \mathbf{k}_2$ close to the Fermi surface being scattering into $\mathbf{k}_3, \mathbf{k}_4$, then $\mathbf{k}_3$ and $\mathbf{k}_4$ can only lie on or near the circle on the Fermi surface where the plane perpendicular to $\mathbf{k}_1 + \mathbf{k}_2$, and which passes through $\mathbf{k}_1$ and $\mathbf{k}_2$, intersects the Fermi sphere. When the scattering is restricted to states on the Fermi surface $\tilde{U}^{\sigma_1\sigma_2}_{\sigma_3\sigma_4}(\mathbf{k}_1, \mathbf{k}_2, \mathbf{k}_3, \mathbf{k}_4)$ can be expressed as $\tilde{U}^{\sigma_1\sigma_2}_{\sigma_3\sigma_4}(\theta, \phi)$ where $\theta$ is the angle between the wavevectors $\mathbf{k}_1$ and $\mathbf{k}_2$ and $\phi$ the angle betwen the plane containing the vectors $(\mathbf{k}_1, \mathbf{k}_2)$ and that containing $(\mathbf{k}_3, \mathbf{k}_4)$. There have been attempts to write down an effective Hamiltonian for interacting quasiparticles with $\mathbf{q}$ dependent scattering and parameters chosen so as to reproduce the Fermi liquid theory for forward scattering (see for instance [35]). However we can see from the theory developed here the limitations and inconsistencies of any such approach which does not take account of the full Hamiltonian and include the counter terms.

(ii). We conclude also that $H_{\mathrm{qp}}$ given by (85) is the Hamiltonian for the Fermi liquid fixed point, plus leading corrections, that we would obtain if we could carry out fully the sequence of renormalization group transformations to remove the higher energy excitations. This is in analogy with the impurity case where we could show this directly. The mean field approximation for this Hamiltonian, like that of the impurity fixed point Hamiltonian, gives the Landau Fermi liquid theory. The molecular field nature of the Landau phenomenological theory has been emphasized in the review of Leggett [19]; here we can give it a more precise interpretation within the framework of the renormalization group.

This form for the fixed point Hamiltonian is in general agreement with the calculations of Shankar [18] based on a perturbative renormalization of diagrams up to one loop level. The leading corrections to the free particle Hamiltonian for translationally invariant systems were found by Shankar to be marginal whereas the leading corrections in the Wilson impurity calculation, discussed in section (2), were the leading order irrelevant ones. This is just a reflection of the fact that for an impurity problem we are interested in the impurities effects which, *after summation*, scale independently of the size of the system, and which therefore behave as $1/V_0$ where $V_0$ is the volume of the system.

(iii). The leading order thermodynamics as $T \to 0$ can be calculated either by normal ordering the quasiparticle Hamiltonian, ignoring the counter terms, and taking the mean field theory, or by self-consistently taking into account the first order Hartree (-Fock) diagram together with the counter term. As in the impurity case the counter terms play no role in the Fermi liquid regime other than subtract off the ground state expectation values. For calculations beyond the very low temperature or low frequency regime the full Hamiltonian (71) with the counter terms must be used. This way of reorganising the perturbation expansion closely parallels the renormalized prescription used in perturbative field theory. Here, the approach is rather more similar to that for the $\phi^4$ field theory than quantum electrodynamics because we do not have the problems associated with gauge fields. In the field theory case there is a choice about which point to make the renormalization, and this freedom is exploited in setting up the Callen-Symanzik renormalization group equations. This freedom does not seem to extend to the situation considered here because, if we expand about any point other than one on the Fermi surface, the $z$ factor will in general become complex, this would seem to preclude the setting up Callen-Symanzik type of renormalization group equations in this context.

This reformulation of Fermi liquid theory as a renormalized perturbation theory links the renormalization group, microscopic Fermi liquid theory, and the phenomenological theory of Landau within a single theoretical framework. There should be scope for its further development and application to a range of physical systems.

# 5  Breakdown of Fermi Liquid Theory

Most metallic systems are not Fermi liquids at very low temperatures because at some critical temperature they undergo a phase transition to a magnetic, superconducting, or some other form of ordered state. Nevertheless if this transition occurs at very low temperatures it can usually be analyzed as an instability within a Fermi liquid, such as in the BCS theory of superconductivity. In the case of superconductivity the quasiparticle scattering becomes singular at the critical temperature $T_c$ signalling the on-set of the superconductivity if there is an attractive interaction between the quasiparticles in one of the particle-particle scattering channels, no matter how small. More commonly, as for most magnetic transitions, there is some critical value of the interaction strength between the quasiparticles, which has to be exceeded if the scattering between the quasiparticles becomes singular. If quasiparticles are assumed to exist in one dimension the scattering is singular whatever the magnitude (non-zero) or sign of the interactions and there is no (finite) critical temperature, indicating the breakdown of the quasiparticle concept in this case [36]. As mentioned earlier Haldane has shown that for one dimensional systems with repulsive interactions the behaviour at low temperatures can be described within the framework of a Luttinger model [14]. This would seem to preclude any description in terms of the renormalized perturbation theory that we introduced in the earlier sections as we assumed the self-energy to be analytic at $\omega=0$ to give a finite wave-function renormalization factor. However, we show here that this problem can be circumvented, at least for the spinless Luttinger model, by making our expansion at finite $\omega$ and combining it with a form of poor man's renormalization. The correct form for the Green's function is obtained with the exponents of the singular terms evaluated perturbationally. A form of renormalized perturbation theory with counter terms can then be set up for the regular contributions to the self-energy. We give an outline of the approach.

In the Luttinger model there are two branches of the conduction states, $a$ and $b$, with Fermi points at $k_F$ and $-k_F$, and energies relative to the Fermi energy of $v_F(k-k_F)$ and $v_F(-k-k_F)$, respectively. The Hamiltonian of the model [10] is

$$H = \sum_k v_F(k-k_F)c^\dagger_{a,k}c_{a,k} + \sum_k v_F(-k-k_F)c^\dagger_{b,k}c_{b,k} + \frac{1}{2}\sum_{k,q,i,j=a,b} v_{i,j}(q)c^\dagger_{i,k-q}c_{j,k'+q}c^\dagger_{j,k'}c_{i,k}, \tag{88}$$

where $v_{i,j}(q)$ is the interaction which is symmetric with respect to $i$ and $j$. In this version of the model the number of electrons of each type are conserved. We take a cut-off of $D$ on the momentum transfer $q$, $|q|<D$ and a cut-off on the bands of $W$ so that $|k-k_F|<W$ for the $a$ band and $|k+k_F|<W$ for the $b$ band with $W>D$. The Luttinger model is an approximation for a one dimensional system of interacting fermions in which the dispersion of $\epsilon(\mathbf{k})$ is linearized about the two points of the Fermi 'surface'.

The corresponding Green's functions can be written in the form,

$$G_i(\omega, k_i) = \frac{1}{\omega - v_F k_i - \Sigma_i(\omega, k_i)}, \tag{89}$$

where $k_a = k - k_F$ and $k_b = -k - k_F$.

We now reduce the cut-off $D$ by $\delta D$ and separate the self-energy into two parts, $\Sigma_i(\omega, k_i) = \Sigma_i^{(\mathrm{rem})}(\omega, k_i) + \Sigma_i^{(\delta D)}(\omega, k_i)$, where $\Sigma_i^{(\delta D)}(\omega, k_i)$ is the contribution from terms in which there is any scattering from the band edge region and $\Sigma_i^{(\mathrm{rem})}(\omega, k_i)$ is the remaining part. For the self-energy $\Sigma_i^{(\delta D)}(\omega, k_i)$ we expand in powers of $(\omega-k_i)/D$ and keep the leading terms. This should be sufficient for calculating the behaviour in the region $(\omega-k_i)\ll D$. If $\Sigma_i^{(\delta D)}(\omega, k_i) = a_i(\delta D) + b_i(\delta D)(\omega-k_i) + O(\omega-k_i)^2$ then for this region we can write

$$G_i(\omega, k_i) = \frac{z_i(\delta D)}{\omega - v_F k_i - \tilde{\Sigma}_i(\omega, k_i)} \tag{90}$$

where $z_i(\delta D) = 1/(1-b_i(\delta D))$, $\tilde{\Sigma}_i(\omega, k_i) = z_i(\delta D)\Sigma_i^{(\mathrm{rem})}(\omega, k_i)$ and $a_i(\delta D)$ is absorbed a a renormalization of $k_F$ and the Fermi velocity $v_F$. Initially we calculate $\Sigma_i^{(\delta D)}(\omega, k_i)$ to second order in perturbation theory and justify this later. We do not attempt a complete solution at this stage but just try to pick up the most important contributions. For the system with a reduced cut-off we can set up a new perturbation expansion for the renormalized self-energy $\tilde{\Sigma}_i(\omega, k_i)$ using the renormalized parameters. This can be continued as long as all the steps in the calculation are valid so generating a set of renormalization group equations. The first order perturbation terms only renormalize the chemical potential so we concentrate on the second order terms for the a branch, which corresponds to figure 6(b), taking $v_{i,i}(q) = 0$ so that the a propagator scatters only with a particle-hole pair in the b branch. This gives

$$\Sigma_a^{(2)}(\omega_n, k_a) = \frac{1}{v_F^2} \int \frac{dk'}{2\pi} \int \frac{v_{a,b}^2(q)dq}{2\pi} \frac{f((k_b' - q)) - f((k_b'))}{(i\omega_n + 2q - k_a)} (n(q) + f((k_a - q))), \quad (91)$$

where we have now scaled the $k$s to absorb the Fermi velocity $v_F$, and $n(q)$ is the Bose distribution function. We integrate over $k'$ for $T = 0$ to obtain

$$\Sigma_a^{(2)}(\omega_n, k_a) = \frac{1}{(2\pi v_F)^2} \int v_{a,b}^2(q) q \, dq \frac{(\theta(q - k_a)\theta(q) - \theta(-q + k_a)\theta(-q)}{(i\omega_n + 2q - k_a)}. \quad (92)$$

We take account of the momentum exchange terms that fall within the band edge. It helps at this stage if we distinguish the upper and lower band edges taking the running values to be $\tilde{D}_1$ and $-\tilde{D}_2$. The second order contribution due to scattering in the band edges is given by

$$2 \left\{ \tilde{\gamma}_1 \delta \tilde{D}_1 - \tilde{\gamma}_2 \delta \tilde{D}_2 - \frac{\tilde{\gamma}_1(\omega - k_a)\delta \tilde{D}_1}{(\omega + k_a) + 2\tilde{D}_1} - \frac{\tilde{\gamma}_2(\omega - k_a)\delta \tilde{D}_2}{2\tilde{D}_2 - (\omega - k_a)} \right\} \quad (93)$$

where $\tilde{\gamma}_i = v_{a,b}^2(\tilde{D}_i)/4(2\pi v_F)^2$ and we have specialized to the case $k_a > 0$ and continued to real frequencies. To eliminate a step function we have also displaced $\tilde{D}_1$ by the transformation to $\tilde{D}_1 - k_a \to \tilde{D}_1$. We can set up scaling equations by expanding to order $\delta D/D$, the most important one being that for the wavefunction renormalization factor $z_a$ which is given by

$$\ln z_a = - \int_{\alpha|\omega+k_a|}^{D} \tilde{\gamma}_1(\tilde{D}_1) \frac{d\tilde{D}_1}{\tilde{D}_1} - \int_{\alpha|\omega-k_a|}^{D} \tilde{\gamma}_2(\tilde{D}_2) \frac{d\tilde{D}_2}{\tilde{D}_2}, \quad (94)$$

where $\alpha$ is a suitable constant subject only to the condition $\alpha \gg 1$. Due to the different regimes of validity of the expansions used the lower limits differ in the two cases.

There is also the possibility of a renormalization of the running effective interaction $\tilde{\gamma}_i$ to consider. There are two second order corrections for the a part of the interaction vertex for the simple model with scattering between the two branches only; one arising from the diagram in figure 9 and the other from the $z_a$ factor which in the renormalized expansion is absorbed into the vertex, and similarly for the b part of the vertex. It is straightforward to show that these two corrections cancel, and so if we take $v_{a,b}(q)$ independent of $q$ then $\tilde{\gamma}_i$ is a constant $\tilde{\gamma}_i = \gamma = v_{a,b}^2/4(2\pi v_F)^2$. This result can also be derived from a Ward identity, based on the conservation of the two species of fermions by the Hamiltonian [37].

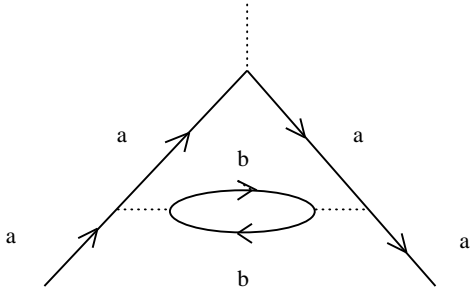

Figure 9. The second order contribution to the renormalized $a$ vertex for the Luttinger model with the momentum exchange $q$ to the $b$ particle-hole pair in the interval $D - \delta D < |q| < D$ .

As $\tilde{\gamma}$ is constant we can integrate (94) which gives the result,

$$z_a = \frac{\alpha^{2\gamma}|\omega + k_a|^\gamma |\omega - k_a|^\gamma}{D^{2\gamma}}. \tag{95}$$

The renormalized self-energy $\Sigma_a^{(2)}(\omega, k_a)$ for remaining the range of $q$, values in the range $-\alpha|\omega - k_a| < q < \alpha|\omega + k_a|$, is calculated. This is not singular at $\omega = k_a$ because of the $\omega$ dependent upper and lower cut-offs. The final result is

$$G_a(\omega, k_a) = \frac{|\omega + k_a|^\gamma |\omega - k_a|^\gamma}{\bar{D}^{2\gamma}(\omega - k_a)}, \tag{96}$$

where $\bar{D}$ is independent of $\alpha$ for $\alpha \gg 1$ to leading order in $\gamma$. The renormalizations of $k_a$ have been absorbed into a renormalized Fermi velocity. This result corresponds to the exact form for the spinless Luttinger model, with the exponent $\gamma$ given to leading order. There is no quasiparticle excitation at $\omega = k_a$ as $z_a$ goes to zero at this point. In this derivation we do not follow the usual scaling procedure of changing the cut-offs on all the momenta. We reduce the cut-off on the momentum transfer $q$ only. In progressively reducing this cut-off we take into account the short range parts of the interaction at each renormalization step.

The second order terms neglected in setting up the scaling equations can be taken in to account by writing the self-energy as the sum, $\Sigma_a(\omega, k_a) = \Sigma_a^{(\text{reg})}(\omega, k_a) + \Sigma_a^{(\text{sing})}(\omega, k_a)$, where is the singular $\Sigma_a^{(\text{sing})}(\omega, k_a)$ is the singular contribution we have just calculated and $\Sigma_a^{(\text{reg})}(\omega, k_a)$ is the remaining regular part. Then we can apply standard perturbation theory for $\Sigma_a(\omega, k_a)$ and deduce $\Sigma_a^{(\text{reg})})(\omega_n, k_a)$ exactly to second order from

$$\Sigma_a^{(\text{reg})}(\omega, k_a) = \Sigma_a(\omega, k_a) - (\tilde{k}_1 - k_a) + (\omega - \tilde{k}_a)(e^{-\gamma \ln(\omega^2 - \tilde{k}_a^2)/\bar{D}^2} - 1) \tag{97}$$

The terms arising from $\Sigma_a^{(\text{sing})}(\omega_n, k_a)$ cancel the non-analytic terms in $\Sigma_a(\omega, k_a)$ to second order $v_{a,b}$ and so regularize the expansion to this order. There are different possible procedures for calculating higher order corrections depending on the range of $\omega$ of primary interest. If the main interest is in the region near $\omega = k_a$ then by introducing an effective coupling $\tilde{v}_{a,b}$ by $\tilde{v}_{a,b}^2 = 4\gamma(2\pi v_{\rm F}^2)$, and compensating counter terms, one can fix the exponent $\gamma$ and calculate the regular terms in powers of $\gamma$. The counter terms are then determined by the condition that they eliminate any further the singular terms in (97) order by order in $\gamma$. We can thus obtain the essential features for the Green's function for the spinless Luttinger model from this simple renormalization group approach. The Green's function for the Luttinger model can be obtained exactly from a complete summation of diagrams [39][40] or via bosonization [41][38]. Here, however, we are concerned with finding approximate techniques which can capture the essential physics of the model that can be generalized to tackle some of the interesting physical models which are not tractable by the exact approaches. An example in this context where this approach might be applicable is the model of two directly coupled Luttinger liquids [42] which has been put forward to explain the enhanced superconductivity of the high $T_{\rm c}$ compounds [43][44].

The two dimensional Hubbard model is the most basic model which has been put forward to describe the electrons in the $CuO_2$ planes of the high $T_{\rm c}$ materials. This model has the form,

$$H = \sum_{\langle ij \rangle \sigma} t_{i,j} c_{i,\sigma}^\dagger c_{j,\sigma} + U \sum_i n_{i\uparrow} n_{i\downarrow}, \qquad (98)$$

where $t_{i,j}$ is the hopping matrix element for electrons in the hybridized orbitals built up from the copper $3{\rm d}(x^2 - y^2)$ and the oxygen p$\sigma$ state, with $U$ as a short range repulsive interaction. Bethe ansatz [45] and conformal invariance methods [46][47] provide a comprehensive picture for the one dimensional version of this model but so far there has been limited progress in understanding the behaviour of the two dimensional model despite much active work in this field. The low energy excitations of the one dimensional model can be analyzed in terms of the holons and spinons of the general Luttinger model (with spin). The low energy features of the one electron Green's function can be obtained from conformal invariance methods [46]. The non-doped high T$_{\rm c}$ materials correspond to the insulating half filled Hubbard model. For an insulating state described by the Hubbard model there must be a Mott-Hubbard gap in the one electron spectrum. In one dimension a gap exists no matter how small the interaction $U$ [45]. In models of higher dimensionality this is only likely to develop at some critical value of $U$ ($= U_{\rm c}$). The superconductivity arises with relatively small doping from this insulating state so it is of some interest to understand the transition from a Fermi liquid to a Mott-Hubbard insulator as $U$ is increased to the critical value $U_{\rm c}$, and also to determine the nature of the low energy excitations away from half-filling for $U > U_{\rm c}$ where the electrons are the most constrained. In this parameter regime, because of the limited phase space available for scattering, the correlation effects are likely to be at their strongest and non-Fermi liquid states might occur or transitions to magnetic order (at half filling and finite but large $U$ the model is equivalent to an antiferromagnetic Heisenberg model). This difficult regime appears to be the most tractable in the infinite dimension version of this model where it has been shown that the self-energy is purely local (independent of the wavevector **k**) and the model can be mapped into an effective impurity model with an additional self-consistency condition [48]. By formally integrating out all the states except those of a single site in a functional integral representation of the partition function the model can be mapped on to an impurity problem with an effective hybridization matrix element. The hybridization can be absorbed into the resonance width function $\Delta(\omega)$ of the Anderson model (10). The Green's function for the non-interacting version of this effective Anderson model can be put in the form,

$$G_{d,\sigma}^{(0)}(\omega) = \frac{1}{\omega - \epsilon_d - 1/\pi \int \Delta(\omega')/(\omega - \omega')d\omega'} \quad (99)$$

which is equivalent to (30). The self-energy $\Sigma(\omega)$ of the impurity model can be formally expressed in the form,

$$\Sigma(\omega) = F(U, G_{d,\sigma}^{(0)}(\omega)), \quad (100)$$

where $F$ is a functional of $G_{d,\sigma}^{(0)}(\omega)$ and corresponds to the complete sum of the diagrammatic perturbation series. This sum is completely specified, given $G_{d,\sigma}^{(0)}(\omega)$ and $U$, and the functional $F$ is universal in the $d \to \infty$ limit. The local or on-site Green's function for the infinite dimensional model has the form,

$$G_l(\omega) = \sum_{\mathbf{k}} \frac{1}{\omega - \epsilon_{\mathbf{k}} - \Sigma(\omega)} = \int \frac{\rho_c(\epsilon)}{\omega - \epsilon - \Sigma(\omega)} d\epsilon, \quad (101)$$

where $\rho_c(\epsilon)$ is the density of states of the non-interacting model (for a tight-binding hypercubic model this has a Gaussian form). The self-consistency condition arises from the fact that this Green's function corresponds to that of the equivalent impurity, and hence has to satisfy the condition,

$$\Sigma(\omega) = G_l^{-1}(\omega) - (G_{d,\sigma}^{(0)}(\omega))^{-1}. \quad (102)$$

The resulting equations are those for the Anderson impurity model but, instead of the resonance width function $\Delta(\omega)$ being specified as an initial condition, it has to be determined self-consistently. As one is now dealing with an impurity model there is a good chance that a reasonably accurate solution can be obtained. The difficult step is to calculate the self-energy given $U$ for any given form of $G_{d,\sigma}^{(0)}(\omega)$. Many methods have been devised for calculating the self-energy and the spectral density for the Anderson model, such a perturbation theory [49], Monte Carlo [50], the non-crossing approximation [51], and the renormalization group [52]. These all have their advantages and disadvantages. The easiest to carry out in practice is the second order perturbation theory which has been found to work very well for the standard impurity model with particle-hole symmetry up to values of $U$ into the strong correlation regime (where there are three well defined peaks in the spectral density corresponding to that shown in figure 5). Second order perturbational results for the local density of states $\rho(\omega)$ of the infinite dimensional model on a Bethe lattice are shown in figure 10 for a value of $U$ greater than then conduction band width but less than the critical value $U_c$. If this is compared with the corresponding density of states for an impurity hybridized with a wide structureless conduction band as shown in figure 5 one sees the tendency of a gap to open up and the central peak associated with the quasiparticle density of states to become isolated. The local density of states for the non-interacting version of this model has the single peaked semi-elliptical form. In the results for the real part of the self-energy shown in figure 11 for the same model one can see a large negative gradient developing at the Fermi level resulting in a small value for the wavefunction renormalization factor $z$. As $U \to U_c$ the self-energy becomes singular at $\omega = 0$ so $z \to 0$ and the quasiparticle peak of the Fermi liquid disappears. This basic picture of the breakdown of the Fermi liquid state is confirmed in the Monte Carlo, non-crossing approximation and renormalization group approaches. There are some complications arising from the fact that there is a another solution corresponding to an insulating state which persists below $U_c$ but this solution appears to be a unstable one for $T = 0$ [53]. With several groups actively working on this problem we might expect to get soon a detailed definitive theory of the breakdown of Fermi liquid theory at the Mott transition for this model.

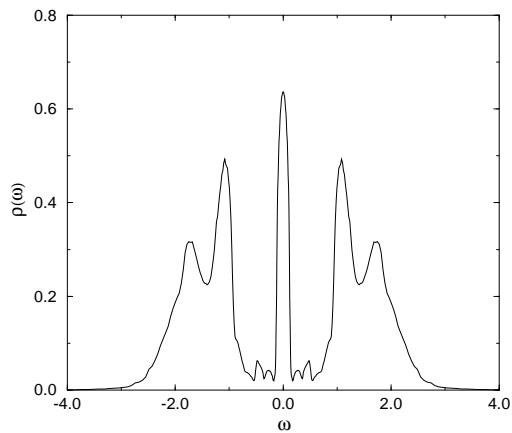

Figure 10. The local density of states for the infinite dimensional Hubbard model at half-filling for a Bethe lattice (total bandwidth $2D = 2.0$ for the non-interacting model) with $U = 2.7$ as calculated from second order perturbation theory for the impurity model with the self-consistency condition (102).

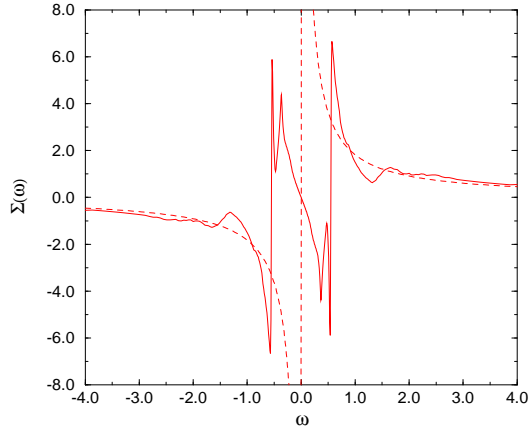

Figure 11. The real part of the self-energy $\Sigma(\omega)$ for the infinite dimensional Hubbard model with the same parameters as given in figure 10. The dashed curve corresponds to the self-energy in the atomic limit.

Breakdowns of Fermi liquid have also been proposed for some strong correlation models away from half-filling and with dimensionality greater than one [4][54]. In this regime the breakdown of Fermi liquid theory is more unusual as the resulting state can be a metal but lacking a Fermi surface in the usual sense as the imaginary part of the self-energy does not vanish, as in the recent theory of Herz and Edwards [4]. There is an impurity model which shows non-Fermi liquid behaviour, the n-channel Kondo impurity model where an impurity spin $S$ is coupled equally to the spins of conduction electrons in each of the $n$ channels. The thermodynamics of this model can be calculated exactly by the Bethe ansatz [55] and non-Fermi liquid behaviour is found in the 'overcompensated' regime $n > 2S$. The low frequency behaviour of the local Green's function can be determined from conformal field theory [57]. This model has been put forward to explain the anomalous behaviour of some uranium heavy fermion materials [56]. The relevance of these various non-Fermi liquid theories to strong correlation systems, and in particular to the high temperature superconductors, is likely to remain a subject of lively debate in the field for sometime. The non-Fermi liquid theories of strong correlation systems is a suitable topic for a future survey.

**Acknowledgement**

I am grateful to the SERC for the support of a research grant.

# References

[1] *Physical Properties of High Temperature Superconductors* ed. D.M. Ginsberg volumes 1-4, 1990-94, World Scientific.

[2] C.M. Varma P.B. Littlewood, S. Schmidt-Rink, E. Abrahams and A.E. Ruckenstein 1989 Phys. Rev. Lett. **63** 1996.

[3] P.W. Anderson, P.W. and Y. Ren, 1990 In *The Los Alamos Symposium 1989, High Temperature Superconductivity Proceedings*, ed. K.S. Bedell, D. Coffey, D.E. Meltzer, D. Pines and J.R. Schrieffer, p. 3, Addison-Wesley.

[4] D.M. Edwards and J.A. Hertz 1990 Physica B **163** 527: D.M. Edwards 1993 J. Phys. Cond. Mat. **5** 161.

[5] L.D. Landau 1957 Sov. Phys. JETP **3** 920 : **5**, 101 ; 1958 Sov. Phys. JETP **8** 70.

[6] J.M. Luttinger 1960 Phys. Rev. **119** 1153; 1961 Phys. Rev. **121** 942.

[7] P. Nozières 1964 *The Theory of Interacting Fermi Systems* Benjamin.

[8] A.A. Abrikosov, L.P. Gorkov and I.E. Dzyaloshinskii 1975 *Methods of Quantum Field Theory in Statistical Physics*, Dover.

[9] S. Tomonaga 1950 Prog. Theor. Phys. **5** 544.

[10] J.M. Luttinger 1963 J. Math. Phys. **4** 1154.

[11] D.C. Mattis and E.H. Lieb 1965 J. Math. Phys. **6** 304.

[12] A. Luther 1979 Phys. Rev. B **19**, 320

[13] V.J. Emery 1979 in *Highly Conducting One Dimensional Solids* Ed. J.T. Devreese, R.P. Evrad, and V. van Doren, Plenum, p 247.

[14] F.D.M. Haldane 1981 J. Phys. C **14**, 2585.

[15] A. Houghton and B. Marston 1993 Phys. Rev. B **48** 7790.

[16] A.H. Castro Neto and E.H. Fradkin 1994 Phys. Rev. B **49** 10877.

[17] K.G. Wilson 1975 Rev. Mod. Phys. **47** 773.

[18] R. Shankar 1994 Rev. Mod. Phys. **66** 129.

[19] A.J. Leggett 1975 Rev. Mod. Phys. **47** 331.

[20] D.Pines and P. Nozières 1966 *The Theory of Quantum Liquids* W.A. Benjamin: G. Baym and C. Pethick 1991 *Landau Fermi-liquid Theory: Concepts and Applications* John Wiley: D. Vollhardt and P. Wölfle 1990 *The Superfluid Phase of Helium 3* Taylor and Francis.

[21] K.G. Wilson, and J. Kogut 1974 Phys. Rep. C **12** 75.

[22] P.W. Anderson 1984 *Basic Notions of Condensed Matter Physics* Benjamin-Cummings.

[23] G. Benfatto and G. Gallivotti 1990 Phys. Rev. B **42** 9967.

[24] H.R. Krishna-murthy, J.W. Wilkins, and K.G. Wilson 1980 Phys. Rev. B **21** 1003 and 1044.

[25] K. Yamada 1975 Prog. Theor. Phys. **53** 970; Prog. Theor. Phys. **54**, 316 ; K. Yosida and K. Yamada 1975 Prog. Theor. Phys. **53** 1286.

[26] A.M. Tsvelick and P.B. Wiegmann 1983 Adv. Phys. **32**: N. Andrei, K. Furuya and J.H. Lowenstein 1983 Rev. Mod. Phys. **55** 331: B. Horvatić and V. Zlatić 1985 J. Physique **46** 145.

[27] P.W. Anderson 1961 Phys. Rev. **124** 41.

[28] A.C. Hewson 1993 J. Phys. Cond. Mat. **5**, 6277.

[29] A.C. Hewson 1993 *The Kondo Problem to Heavy Fermions*. Cambridge University Press.

[30] P. Nozières 1974 J. Low Temp. Phys. **17** 31: 1975 Low Temperature Physics Conference Proceedings LT14 **5**, eds. Krusius and Vuorio. p. 339. North Holland/Elsevier.

[31] T. A. Costi and A. C. Hewson 1992 Phil. Mag. B **65**, 1165 : 1993 J. Phys. Cond. Mat. **30**, L361: T. A. Costi, A.C. Hewson and V. Zlatić 1994 J. Phys. Cond Mat **6** 2519.

[32] A.C. Hewson 1993 Phys. Rev. Lett. **70** 4007.

[33] N.N. Bogoliubov, and D.V. Shirkov, *Introduction to the Theory of Quantised Fields*. (3rd edition) Wiley-Interscience (1980); L.H. Ryder, *Quantum Field Theory*. Cambridge University Press (1985).

[34] J. Friedel 1956 Can. J. Phys. **54** 1190; J.M. Langer and V. Ambegaokar 1961 Phys. Rev. **164** 498; D.C. Langreth 1966 Phys. Rev. **150** 516.

[35] G.D. Mahan 1981 *Many-body Physics* Chapter 10, Plenum.

[36] J. Sólyom 1979 Advances in Physics **28** 201.

[37] C. Di Castro and W. Metzner 1991 Phys. Rev. Lett. **67** 3852.

[38] V. Meden and K. Schönhammer 1992 Phys. Rev. B **46** 15753.

[39] I.E. Dzyaloshinskii and A.I. Larkin 1974 J.E.T.P **38** 202.

[40] K. Penc and J. Sólyom 1991 Phys. Rev. B **44** 12690.

[41] D.K.K. Lee and Y. Chen 1988 J. Phys. A **21** 4155.

[42] C. Di Castro and W. Metzner 1992 Phys. Rev. Lett. **69** 1703.

[43] S. Chakravarty, A. Sudbø, P.W. Anderson and S.P. Strong 1993 Science **261** 337.

[44] D.G. Clarke, S.P. Strong and P.W. Anderson 1994 Phys. Rev. Lett. **72** 3218.

[45] E.H. Lieb and F.Y. Wu 1689 Phys. Rev. Lett. **20** 1447.

[46] H. Frahm and V. Korepin 1990 Phys. Rev. B **42** 10553.

[47] N. Kawakami and S-K Yang 1991 J. Phys. Cond. Mat. **3** 5983.

[48] E. Müller–Hartmann, Z. Phys. **B74** 507 (1989): W. Metzner and D. Vollhardt 1989 Phys. Rev. Lett. **62**, 324: A. Georges and G. Kotliar 1992 Phys. Rev. **B45** 6479 : D.M. Edwards 1993 J. Phys. Cond. Mat. **5** 161: A. Georges and W. Krauth 1993 Phys. Rev. B **48** 7167.

[49] X.Y. Zhang, M.J. Rozenberg and G. Kotliar 1993 Phys. Rev. Lett. **70**, 1666.

[50] M. Jarrell 1992 Phys. Rev. Lett. **69** 168: M. Jarrell, Hossein Akhlaghpour and Th. Pruschke 1993 Phys. Rev. Lett. **70**, 1670.

[51] M. Jarrell and Th. Pruschke 1994 Phys. Rev. B **49** 1458.

[52] O. Sakai and Y. Kuramoto 1994 Sol. St. Com. **89** 307.

[53] M.J. Rozenberg, G. Kotliar and X.Y. Zhang 1994 Preprint.

[54] Q. Si and G. Kotliar 1993 Phys. Rev. Lett **70** 3143.

[55] A.M. Tsvelick and P.B. Wiegmann 1984 Z. Phys. B **54** 201: N. Andrei and C. Destri 1984 Phys. Rev. Lett. **52** 364: P. Schlottmann and P.D. Sacramento 1993 Advances in Physics **42** 641.

[56] D.L. Cox 1977 Phys. Rev. Lett. **59** 1240: 1988 Physica C **153** 1642.

[57] A.W.W. Ludwig and I. Affleck 1991 Phys. Rev. Lett **67** 3160.